\documentclass[sigconf, nonacm]{acmart}

\usepackage{graphicx}
\usepackage{subcaption}
\usepackage{amsmath}
\usepackage{hyperref}
\usepackage{multirow}
\usepackage{xspace}
\usepackage{enumitem}

\newcommand{\sqlshare}{\textsf{SQLShare}\xspace}
\newcommand{\sdss}{\textsf{SDSS}\xspace}
\newcommand{\joinorder}{\textsf{Join-Order}\xspace}
\newcommand{\spider}{\textsf{Spider}\xspace}

\newcommand{\mistralai}{\textsf{MistralAI}\xspace}
\newcommand{\gptthree}{\textsf{GPT3.5}\xspace}
\newcommand{\gptfour}{\textsf{GPT4}\xspace}
\newcommand{\llama}{\textsf{Llama3}\xspace}
\newcommand{\gemini}{\textsf{Gemini}\xspace}

\newcommand{\qlen}{\texttt{char\_count}\xspace}
\newcommand{\qtype}{\texttt{query\_type}\xspace}
\newcommand{\wcount}{\texttt{word\_count}\xspace}
\newcommand{\fcount}{\texttt{function\_count}\xspace}
\newcommand{\pcount}{\texttt{predicate\_count}\xspace}
\newcommand{\tcount}{\texttt{table\_count}\xspace}
\newcommand{\ccount}{\texttt{column\_count}\xspace}
\newcommand{\jcount}{\texttt{join\_count}\xspace}
\newcommand{\nlevel}{\texttt{nestedness}\xspace}
\newcommand{\qaggregate}{\texttt{aggregate}\xspace}

\newcommand{\synerror}{\texttt{syntax\_error}\xspace}
\newcommand{\mtoken}{\texttt{miss\_token}\xspace}
\newcommand{\qequiv}{\texttt{query\_equiv}\xspace}
\newcommand{\synerrort}{\texttt{syntax\_error\_type}\xspace}
\newcommand{\mtokent}{\texttt{miss\_token\_type}\xspace}
\newcommand{\qequivt}{\texttt{query\_equiv\_type}\xspace}
\newcommand{\qexplain}
{\texttt{query\_exp}\xspace}
\newcommand{\qperformance}
{\texttt{performance\_pred}\xspace}
\newcommand{\mtokenl}{\texttt{miss\_token\_loc}\xspace}

\usepackage{listings}

\definecolor{light-gray}{gray}{0.97}

\lstdefinestyle{promptstyle}{
    basicstyle=\rmfamily\footnotesize,
    backgroundcolor=\color{light-gray},
    keywordstyle=\color{blue},
    stringstyle=\color{red},
    commentstyle=\color{gray},
    showstringspaces=false,
    breaklines=true,
    tabsize=4
}

\newcommand{\mybox}[1]{\raisebox{0.5ex}{\colorbox{#1}{\rule{0pt}{0.1pt}\rule{0.1pt}{0pt}}}}

\definecolor{myblue}{RGB}{0, 114, 178}
\definecolor{myorange}{RGB}{230, 159, 0}
\definecolor{mygreen}{RGB}{3, 135, 9}
\definecolor{mypink}{RGB}{204, 121, 167}
\definecolor{mybrown}{RGB}{121, 79, 48}
\definecolor{myyellow}{RGB}{240, 228, 66}

\newcommand{\takeaway}[1]{\vspace{1mm}\noindent \uline{Takeaways:} #1}
\newcommand{\ignore}[1]{}
\newcommand{\mostafa}[1]{\noindent \textcolor{red}{{\bf Mostafa:} #1}}
\newcommand{\paraStart}[1]{\vspace{1mm} \noindent {{\bf #1}}}

\newcommand{\paraUStart}[1]{\noindent {\uline{#1}}}

\newcommand{\rachel}[1]{\noindent \textcolor{purple}{{\bf Rachel:} #1}}

\newcommand{\feic}[1]{\noindent \textcolor{blue}{{\bf FeiComment:} #1}}

\newcommand{\ananya}[1]{\noindent \textcolor{green}{{\bf Ananya:} #1}}

\newcommand{\TP}{TP\xspace}
\newcommand{\TN}{TN\xspace}
\newcommand{\FP}{FP\xspace}
\newcommand{\FN}{FN\xspace}
\usepackage[normalem]{ulem}

\newif\ifshowstd
\showstdfalse 

\newcommand{\accstd}[2]{#1\ifshowstd~$\pm$~#2\fi}
\newcommand{\eat}[1]{}

\newcommand\vldbdoi{XX.XX/XXX.XX}

\newcommand\vldbavailabilityurl{URL_TO_YOUR_ARTIFACTS}

\begin{document}
\title{[Experiments \& Analysis] Evaluating SQL Understanding in Large Language Models}

\author{Ananya Rahaman}
\affiliation{%
  \institution{University of Western Ontario}
  \city{London}
  \state{Ontario}
}
\email{arahaman@uwo.ca}

\author{Anny Zheng}
\affiliation{%
  \institution{University of Western Ontario}
  \city{London}
  \state{Ontario}
}
\email{azheng45@uwo.ca}

\author{Mostafa Milani}
\affiliation{%
  \institution{University of Western Ontario}
  \city{London}
  \state{Ontario}
}
\email{mostafa.milani@uwo.ca}

\author{Fei Chiang}
\affiliation{%
  \institution{McMaster Unviersity}
  \city{Hamilton}
  \state{Ontario}
}
\email{fchiang@mcmaster.ca}

\author{Rachel Pottinger}
\affiliation{%
  \institution{University of British Columbia}
  \city{Vancouver}
  \state{British Columbia}
}
\email{rap@cs.ubc.ca}

\begin{abstract}
The rise of large language models (LLMs) has significantly impacted various domains, including natural language processing (NLP) and image generation, by making complex computational tasks more accessible. While LLMs demonstrate impressive generative capabilities, there is an ongoing debate about their level of ``understanding,'' particularly in structured domains like SQL. In this paper, we evaluate the extent to which LLMs ``understand'' SQL by testing them on a series of key SQL tasks. These tasks, such as syntax error detection, missing token identification, query performance prediction, query equivalence checking, and query explanation, assess the models' proficiency in recognition, context awareness, semantics, and coherence—skills essential for SQL understanding. We generate labeled datasets from well-known workloads, and evaluate the latest LLMs, focusing on how query complexity and syntactic features influence performance. Our results indicate that while \gptfour excels at tasks requiring recognition and context, all models struggle with deeper semantic understanding and coherence, especially in query equivalence and performance estimation, revealing the limitations of current LLMs in achieving full SQL comprehension.
\end{abstract}

\maketitle




\section{Introduction}
\label{sec:intro}
The rise of LLMs is having a significant impact across all domains, making computational and data science tasks more accessible and efficient.  For example, in areas such as NLP and image generation, LLMs are able to generate human-like text and realistic images. While LLMs clearly do not have the same level of ``understanding'' as humans, their ability to solve problems (for which they are not directly trained) has alluded  to some degree of ``understanding''~\cite{truong2023language,lee2024tasks,qianlimitations}. Thus, for LLMs, ``understanding'' refers to the model’s ability to perform fundamental tasks at least as proficiently as humans, and potentially even better, across different contexts.  

This level of proficiency can be measured against a set of characteristic \emph{skills} to assess understanding. \emph{Recognition} involves identifying the intended object/entity of interest, e.g., identifying and differentiating between objects and scenes in image generation. \emph{Semantics} involves identifying how meaning is constructed and interpreted, e.g., the meaning of a red octagon is to stop, grasping the meaning of words and phrases.   \emph{Context} defines the scope and setting in which the semantics are interpreted, e.g., in NLP, resolving ambiguities when there exist multiple meanings based on context, comprehending intent (understanding the purpose behind a speaker's words, such as detecting sarcasm or politeness), and handling out-of-distribution elements (identifying when an object or scenario doesn't fit familiar patterns).  Lastly, \emph{coherence} identifies the logical interconnection between objects, e.g., object coherence ensures that objects are placed in realistic positions relative to each other in image generation and identifies the logical links between words, sentences, and paragraphs sharing the same meaning.  Achieving ``understanding'' requires models to demonstrate (increasing) proficiency in these skills and to complete task-specific operations accurately and meaningfully.  Developing a deeper insight into LLM's ``understanding'' is crucial for reliable performance in real-world applications where accuracy is essential.

Toward this goal, we study how LLMs can be used in data management applications, particularly, their ability to perform SQL-related tasks.  Our study goes beyond simply content generation; it evaluates specific SQL tasks that exhibit the aforementioned skills.  We pose the question: \emph{How well do LLMs ``understand" SQL?} 


\paraStart{SQL Tasks.} We propose a series of core SQL tasks designed to probe the depth of LLMs’ SQL ``understanding". \eat{These tasks are important in many applications related to SQL, and we have designed them to progress from simpler tasks, such as syntax error detection, to more complex ones, such as query equivalence checking and explanation.} Novice to advanced SQL users perform tasks ranging from syntax error identification to query performance estimation to query equivalence and explanation.  We evaluate LLMs' ability to perform such tasks, in increasing order of difficulty to reflect increasing skill proficiency.  



\paraUStart{Syntax error identification.} Detecting advanced syntax errors that violate structural and semantic requirements vs. basic errors (e.g., missing parentheses) reflect varying levels of SQL ``understanding". For example, detecting the misalignment of attributes, aggregation functions among \texttt{SELECT}, \texttt{GROUP BY}, \texttt{HAVING} clauses\eat{incorrect use of aggregation functions, such as misapplying the \texttt{GROUP BY} or \texttt{HAVING} clause}, incompatible attribute types between outer and inner queries, and invalid join operations require a complex ``understanding'' of the queries.

\paraUStart{Missing token identification.}
Identifying missing tokens is a crucial pre-step for applications such as query recommendation, where missing token imputation and query auto-completion are key functionalities~\cite{lai2023workload,zolaktaf2020facilitating}. We evaluate the ability to not only recognize a missing token but to identify the precise location and the type of missing token (e.g., missing keywords (e.g., \texttt{SELECT} or \texttt{WHERE}), table names, aliases used in joins or conditions, or literal values.


\paraUStart{Query performance estimation.}  Given only the SQL query text, accurately estimating its runtime performance is challenging, as multiple factors such as the database schema, specific data instances, and the query workload all play a role~\cite{zolaktaf2020facilitating}.  Using publicly available query workloads, recent work has shown that more complex, longer queries with multiple joins and multiple predicate conditions incur higher execution costs~\cite{SDSS,SQLShare,JOB}\eat{\feic{Other references?}}.  We evaluate LLM ``understanding" of query complexity for performance estimation, going beyond surface-level syntax.


\paraUStart{Query equivalence.} Two syntactically different queries are equivalent if they return the same result for all database instances. This is important for query optimization~\cite{chaudhuri1998overview,kim1982optimizing}, and query recommendation~\cite{zolaktaf2020facilitating}, where simpler query representations facilitate faster execution times. We evaluate query equivalence using labeled, equivalent (positive), and non-equivalent (negative) query pairs. While generating equivalent pairs is subtle, the negative case requires careful consideration. If we pair random, non-equivalent queries and label them as such, then the task becomes overly simplistic, as superficial differences will often identify non-equivalence, without testing the model’s ability to understand deeper query semantics.



\eat{We investigate whether LLMs can identify SQL query equivalence, where two queries are considered equivalent if they return the same results for all database instances, even though they may be written differently. This task is essential for various database operations, including The task requires labeled data, consisting of pairs of equivalent queries as well as non-equivalent query pairs, where determining non-equivalence is not straightforward.
}

\paraUStart{Query explainability.} We evaluate LLMs to explain SQL queries by describing the query output. This task is similar to assessments in code and image understanding, to generate code documentation~\cite{nam2024using} and image captions, respectively, to measure understanding. We evaluate over a  wide range of complex queries, including multiple tables, nested subqueries, and intricate logical conditions.

\begin{table}[t!]
    \centering
\small{
\begin{tabular}{l|ccccc}
\toprule
Skill & syntax & missing & Q.perf. &	Q.equiv. & Q.explain.	\cr
 & error  & token & estimate & 	 & 	 \cr
\midrule
Recognition		& $\pmb{\checkmark}$	& $\pmb{\checkmark}$	&  	&	&	\cr
Semantics		& & 	&  	& $\pmb{\checkmark}$	& $\pmb{\checkmark}$	\cr
Context		& &  $\pmb{\checkmark}$ 	& $\pmb{\checkmark}$ 	&	&	 $\pmb{\checkmark}$ \cr
Coherence		& $\pmb{\checkmark}$ & 	& $\pmb{\checkmark}$ 	& $\pmb{\checkmark}$	&	\cr
\bottomrule
\end{tabular}
    \caption{Skill-to-SQL task mapping}
    \label{tab:sqlskills}
\vspace{-8ex}
}
\end{table}

\paraStart{Skill to task proficiency.} 
Each SQL task is associated to a set of skills to assess (SQL) understanding, as summarized in Table~\ref{tab:sqlskills}. \eat{directly to the skills that represent understanding in LLMs, as outlined earlier.} In syntax error identification, we assess recognition and coherence to\eat{relates to both recognition and coherence, as the model} identify syntactic violations, and ensure logical consistency between clauses, such as mismatches between aggregation functions and \texttt{GROUP BY} clauses. Missing token identification requires recognition and context, to detect missing tokens and to determine the correct type and location (e.g., keyword, table name, or value). Query performance estimation evaluates context and coherence, as the model considers the query complexity, the database schema, and the workload to estimate performance. In query equivalence, we assess the model's semantics and coherence abilities to interpret different syntactic query formats but representing the same functional output. \eat{recognize when two syntactically different queries produce the same result.} Lastly, in query explanation, we evaluate the model's semantics and context to describe the query's purpose, and it's result within the context of the schema and the data.

\paraStart{Paper Contributions.} We present an experimental study evaluating the performance of the major LLMs over core SQL tasks. 

\paraUStart{SQL task-driven data benchmark.}
Many of these tasks require labeled data, which we generate by modifying raw queries from popular SQL workloads, such as the Sloan Digital Sky Survey (\sdss)~\cite{SDSS}, SQLShare~\cite{SQLShare}, and Join Order~\cite{JOB}. For syntax error and missing token identification tasks, we create semi-synthetic datasets by randomly selecting queries from workloads and injecting errors or by removing tokens. For each task, we select an appropriate type, such as the type of syntax error to inject or the type of missing token (e.g., keyword, table name, column name). For query performance tasks, we rely on the SDSS workload, which includes log information from past query evaluations. We classify queries based on their runtime, where high runtime represents computationally expensive queries. For query equivalence tasks, we manually modify selected queries to generate equivalent and non-equivalent pairs, ensuring that the modifications reflect realistic query transformations, such as rewriting nested queries using joins.  Our SQL task-driven data benchmark is publicly available.\footnote{\url{https://github.com/AnanyaRahaman/LLMs_SQL_Understading}} \eat{\feic{include link.}}

\paraUStart{Prompt-to-SQL task performance.} Prompt tunning is key in ensuring consistent results from LLMs. We experiment with various prompts, testing them in small-scale trials using a subset of labeled data to identify the best prompt per task. However, interaction with LLMs goes beyond prompt design. Processing their responses is complex, and in our work, we addressed this by using a combination of automated scripts and manual checks to extract the labels.

\paraUStart{SQL task evaluation framework.}
We systematically evaluate the factors influencing LLM performance across SQL tasks.  Our evaluation framework considers three key dimensions. First, we compare SQL task performance across different LLMs. Second, we analyze the properties of the query workloads, particularly the syntactic complexity of the SQL queries, such as the number of tables, conditions, nested subqueries, and overall query length.  We investigate how these syntactic properties affect the LLMs' ability to process and understand queries. Finally, we evaluate the performance of specific SQL tasks across varying parameters, e.g., \eat{the LLMs are asked to perform, such as detecting syntax errors or determining query equivalence. This includes examining} how different types of missing tokens or query transformations affect LLM performance, and whether certain forms of query equivalence or error detection are more challenging to recognize. By considering these three dimensions: LLM performance comparison, workload properties, and task types, we aim to provide a comprehensive evaluation of the factors that influence LLM performance in SQL tasks.

\paraUStart{Extensive comparative evaluation.}
Our experiments show that \gptfour performs best across most tasks, while no other model consistently ranks second. Although most models demonstrate strong performance in binary class tasks such as identifying syntax errors or missing tokens, all LLMs face challenges and suffer reduced accuracy in multi-class tasks,  such as identifying the type of missing token or syntax error. LLMs generally struggle with longer and more complex queries, particularly involving logical reasoning or numerical computations, consistent with prior results~\cite{frieder2024mathematical,chowdhery2023palm,wei2022chain}.





\eat{\mostafa{About understanding vs comprehension, I removed the term ``comprehension'' (let me know if any is left) and consistently used ``understanding'' and also clearly explained what we mean by understanding as per Rachel's suggestion. This way, there won't be a loose term, I agree comprehension vs understanding was unclear.}
\rachel{I like both the red paragraph and the other changes. I made a few minor edits along those lines to help make the changes consistent.}}

Our experimental results demonstrate that LLMs perform well at tasks requiring recognition and context, such as syntax error detection and missing token identification. However, for tasks requiring coherence and semantic understanding, such as query equivalence and performance estimation, the models exhibit limitations. This suggests that while LLMs demonstrate proficiency at surface-level ``understanding", they struggle to fully comprehend deeper semantic relationships, and logical coherence in SQL queries, underscoring the need for further improvements.

\noindent \textbf{Paper Organization.} In Section~\ref{sec:workload}, we describe the workloads used in the study, including comprehensive statistics on the syntactic properties of the queries in our datasets. Section~\ref{sec:setup} outlines the experimental setup, covering the SQL \eat{understanding} tasks, data generation from the workloads, the LLMs, and our interaction with the LLMs, such as prompt tuning. Section~\ref{sec:experiments} presents the experimental results and analysis, Section~\ref{sec:related-work} reviews related work, and we conclude the paper and discusses directions for future work in Section~\ref{sec:future}.



\section{Query Workloads}\label{sec:workload}

A query workload, or simply a workload, is a collection of SQL queries executed against a database, used to simulate real-world usage patterns for performance evaluation and optimization. We give an overview and detailed analysis of the four workloads used in our experimental study.

\paraStart{The Sloan Digital Sky Survey (\sdss) dataset~\cite{SDSS}.} \sdss consists of a relational database with data from a major astronomical survey providing detailed images and spectra of the sky and a SQL query workload used to interact with the \sdss database. \eat{The queries in this workload are designed to extract and analyze complex astronomical data.} The \sdss workload is characterized by its complexity and the need for precise astronomical data retrieval. The workload has been collected over two decades, containing millions of queries. In our study, we use queries recorded in 2023. 

\paraStart{\sqlshare~\cite{SQLShare}.} \sqlshare is an open data platform designed to make data sharing and querying more accessible. The \sqlshare workload consists of a diverse set of user-generated SQL queries, ranging from simple data retrieval to complex data manipulation tasks. Unlike our other workloads, \sqlshare consists of query statements over several databases with different schemas.

\paraStart{\joinorder~\cite{JOB}.} The \joinorder Benchmark is a synthetic workload designed to evaluate the performance of database systems in optimizing join queries. The benchmark includes complex SQL queries to test the optimizer's ability to find efficient join orders. 

\paraStart{\spider~\cite{Spider}.} \spider is a large-scale, complex, cross-domain Text-to-SQL benchmark used to evaluate a model's natural language understanding, and SQL generation capabilities. It includes a wide range of databases, to evaluate generalization across different database schemas. Spider is used extensively in NLP to benchmark model performance to translate natural language queries to SQL.  We use the \spider dataset exclusively for query explanation, while the other three workloads are used for the remaining tasks.

\begin{table}[t!]
    \centering
    \resizebox{\linewidth}{!}{
    \begin{tabular}{lcccccccc} 
        \toprule
        \multirow{2}{*}{\textbf{Workload}} & \multicolumn{2}{c}{\textbf{Number of Queries}} & \multicolumn{2}{c}{\textbf{Query Type}} & \multicolumn{2}{c}{\textbf{Aggregate}} & \multicolumn{2}{c}{\textbf{NestLvl}} \\
        \cmidrule(lr){2-3} \cmidrule(lr){4-5} \cmidrule(lr){6-7} \cmidrule(lr){8-9}
         & Original & Sampled & \texttt{SELECT} & \texttt{CREATE} & Yes & No & 0 & 1 \\
        \midrule
        \textbf{SDSS} & 5,081,188 & 285 & \multicolumn{2}{c}{Fig~\ref{fig:qtypesdss}} & \ 21 & \ 264 & \multicolumn{2}{c}{Fig~\ref{fig:nested-sdss}} \\
        \textbf{SQLShare} & 9,623 & 250 & \multicolumn{2}{c}{Fig~\ref{fig:qtypesqlshare}} & \ 59 & \ 192 & \multicolumn{2}{c}{Fig~\ref{fig:nested-sqlshare}} \\
        \textbf{Join-Order} & 157 & 157 & 113 & 44 & \ 119 & \ 38 & - & - \\
        \textbf{Spider} & 4, 486 & 200 & 200 & 0 & 96 & 104 & 185 & 15 \\
        \bottomrule
    \end{tabular}
    }
    \caption{Workload statistics overview}
    \label{tab:dataset_stats}
    \vspace{-6ex}
\end{table}

All our workloads, with the exception of \joinorder, contain a large number of queries;  making it impractical to utilize all queries in our study.  Therefore, we created smaller datasets by sampling a limited number of queries. 
\ignore{\mostafa{I hide it after addressing it in the table.}\rachel{what is the unit for ``size''? number of queries?}}  Table~\ref{tab:dataset_stats} shows the ``original'' number of queries in the workloads, and ``sampled'' shows the sampled number of queries we use in our experiments.  We describe the dataset generation process in Section~\ref{sec:prepare}. \eat{following the discussion on SQL understanding tasks.} Next, we examine the syntactic properties of our sampled queries, to provide context to interpret our experimental results. \eat{which will assist in interpreting the results in later sections.} Henceforth, we will use \eat{Throughout the paper, we also use} \sdss, \sqlshare, \joinorder, and \spider to refer to the datasets created from the sampled queries of the original workloads.

\begin{figure*}[h!]
    \begin{subfigure}{0.18\textwidth}
        \includegraphics[width=\textwidth]{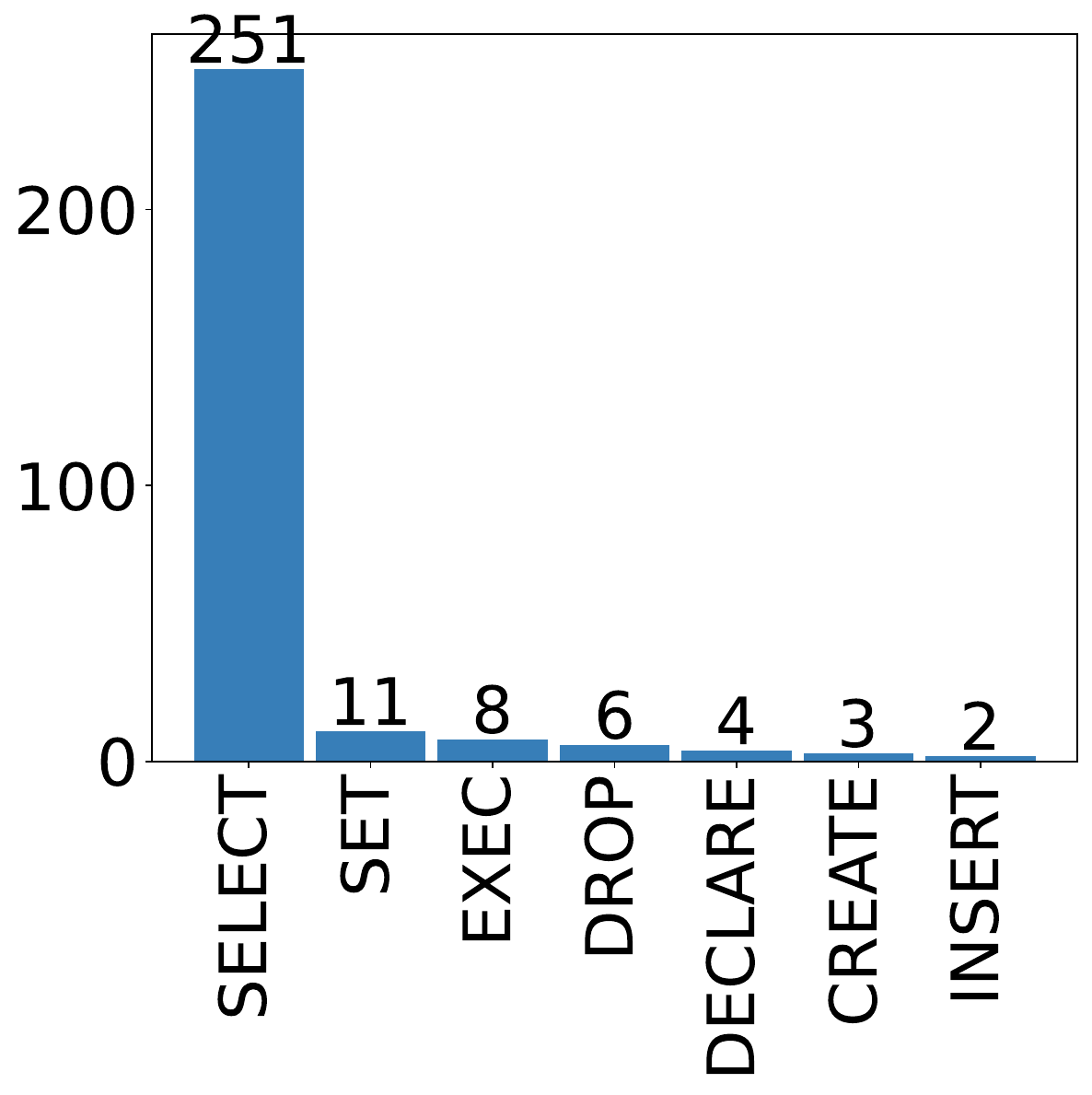}
        \caption{\qtype}
        \label{fig:qtypesdss}
    \end{subfigure}
    \begin{subfigure}{0.18\textwidth}
        \includegraphics[width=\textwidth]{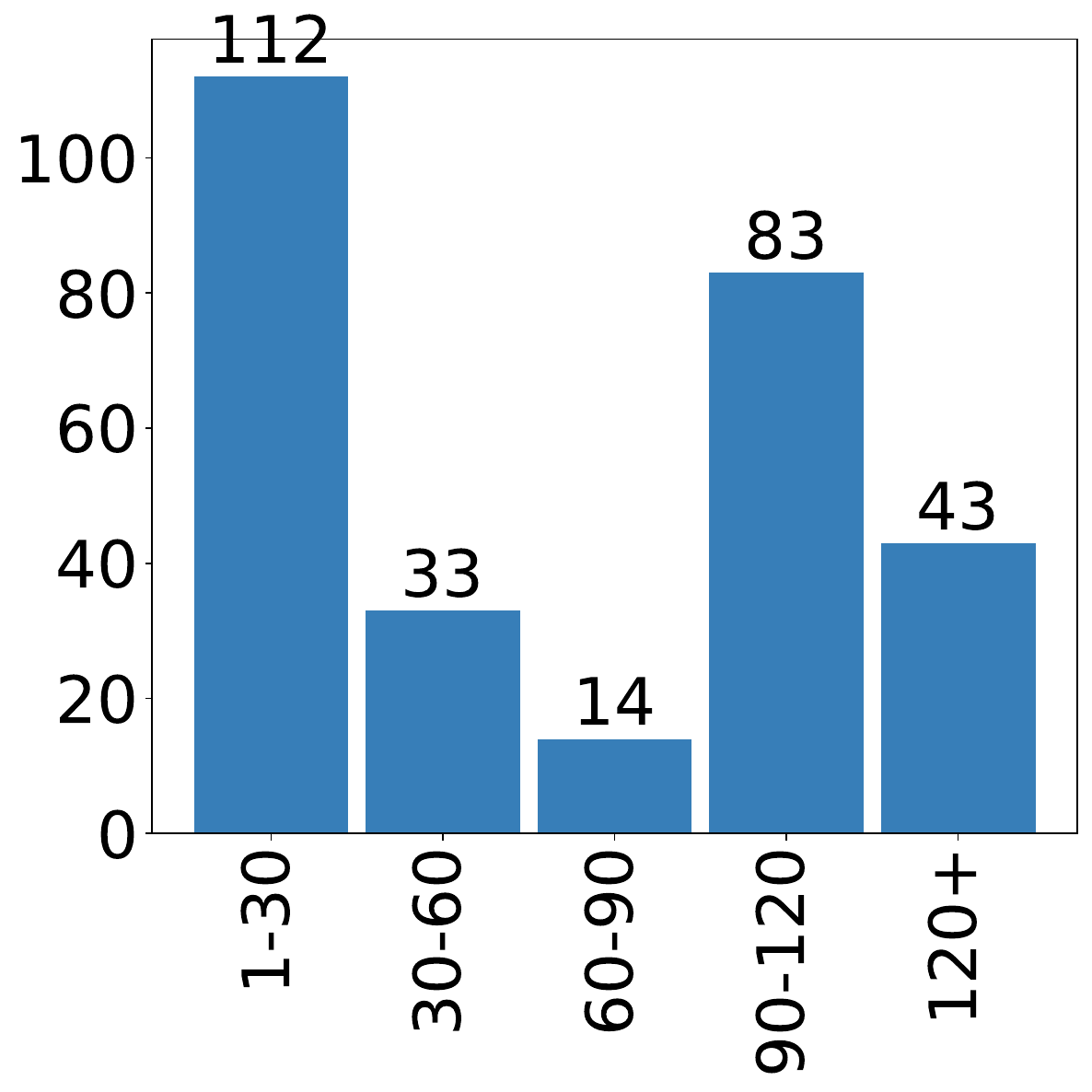}
        \caption{\wcount}
        \label{fig:wcsdss}
    \end{subfigure}
    \begin{subfigure}[b]{0.18\textwidth}
        \includegraphics[width=\textwidth]{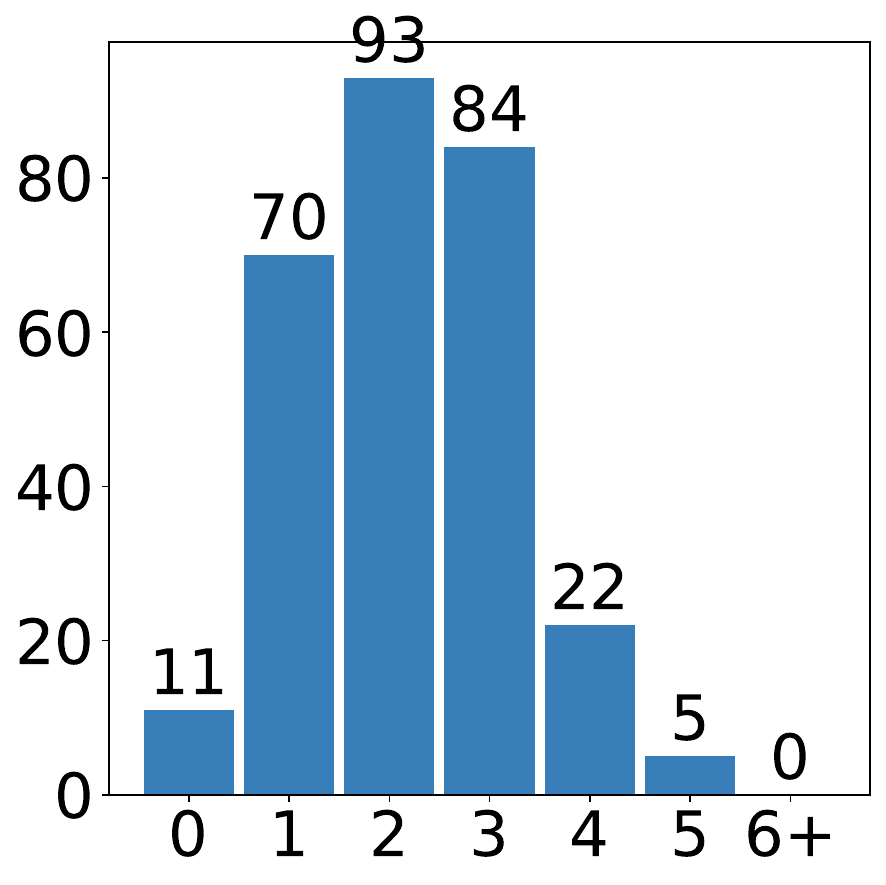}
        \caption{\tcount}
        \label{fig:tcsdss}
    \end{subfigure}
    \begin{subfigure}[b]{0.18\textwidth}
        \includegraphics[width=\textwidth]{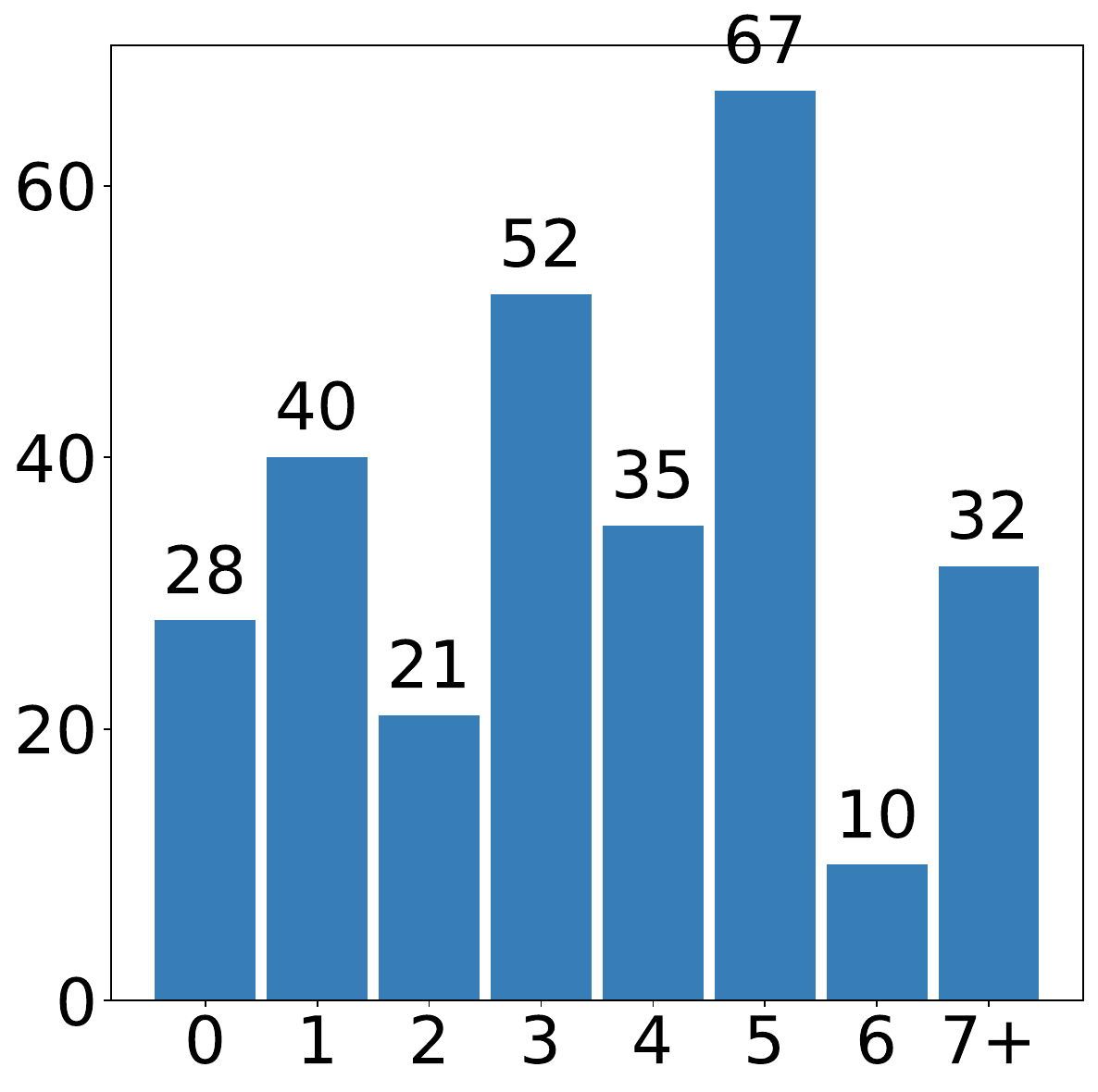}
        \caption{\pcount}
        \label{fig:pcsdss}
    \end{subfigure}
    \begin{subfigure}[b]{0.18\textwidth}
        \includegraphics[width=\textwidth]{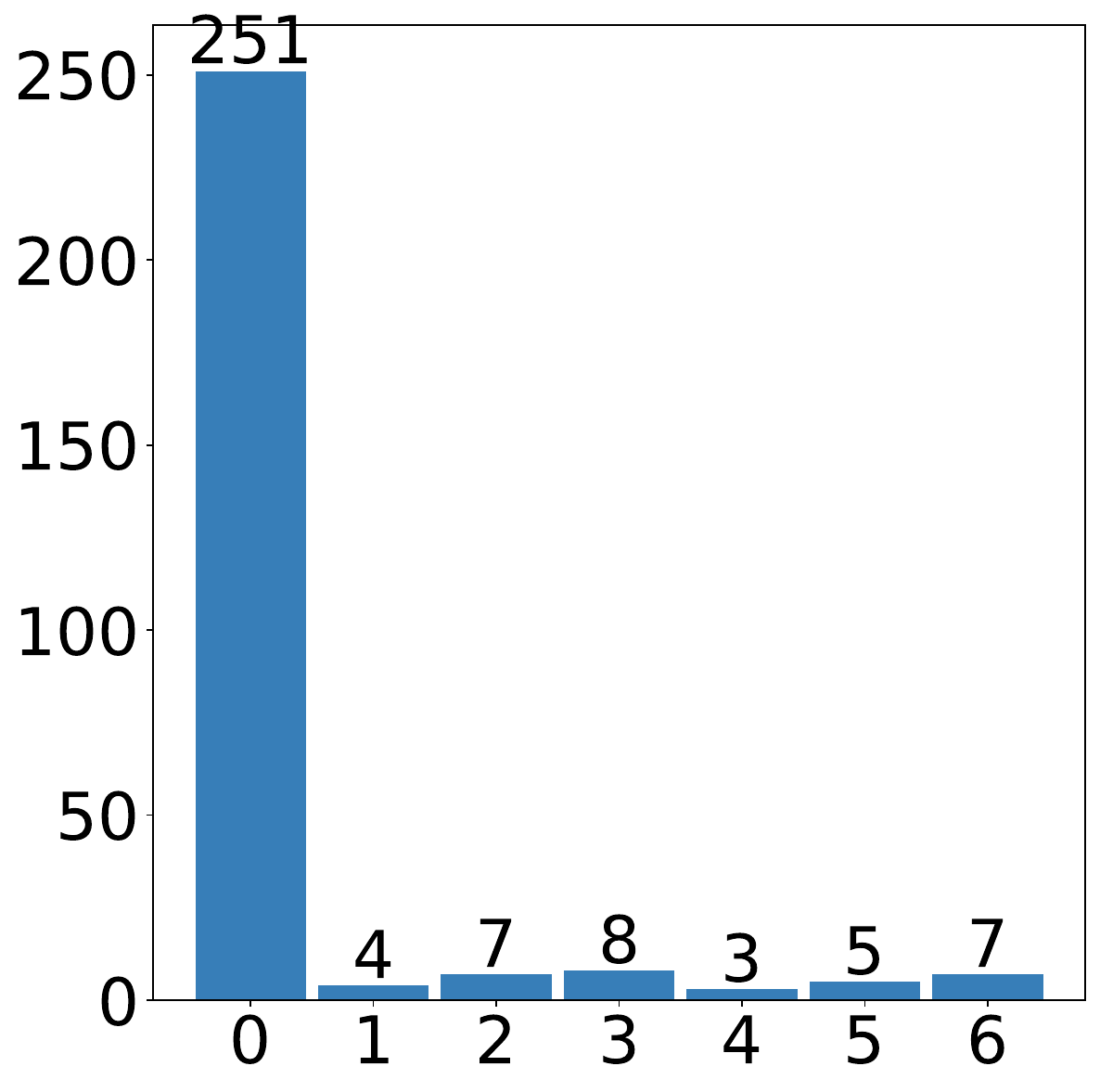}
        \caption{\nlevel}
        \label{fig:nested-sdss}
    \end{subfigure}
    \vspace{-3mm}
    \caption{SDSS Statistics}
    \label{fig:sdss}
\end{figure*}

\begin{figure*}[h!]
    \begin{subfigure}[b]{0.18\textwidth}
        \includegraphics[width=\textwidth]{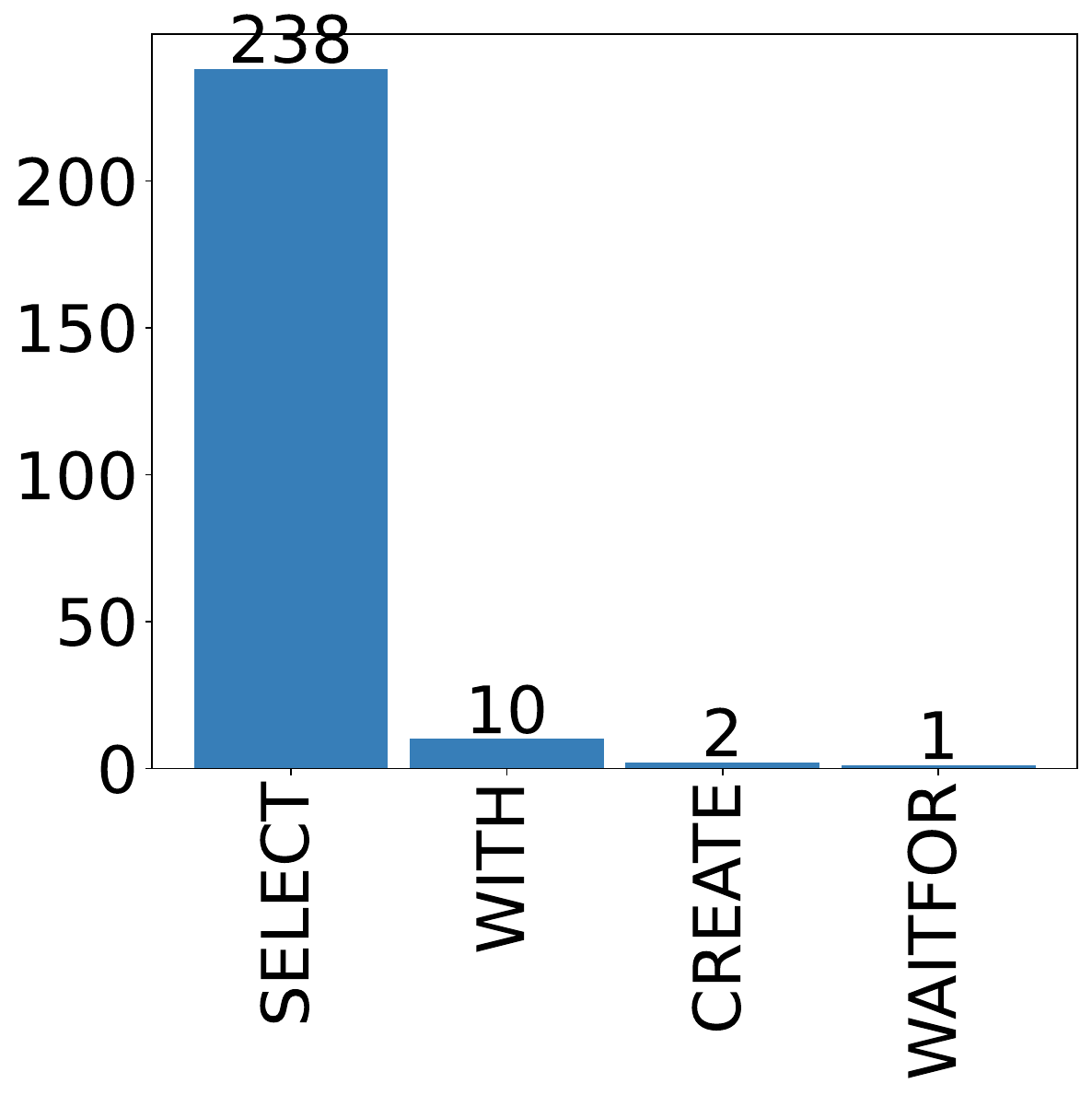}
        \caption{\qtype}
        \label{fig:qtypesqlshare}
    \end{subfigure}
    \begin{subfigure}[b]{0.18\textwidth}
        \includegraphics[width=\textwidth]{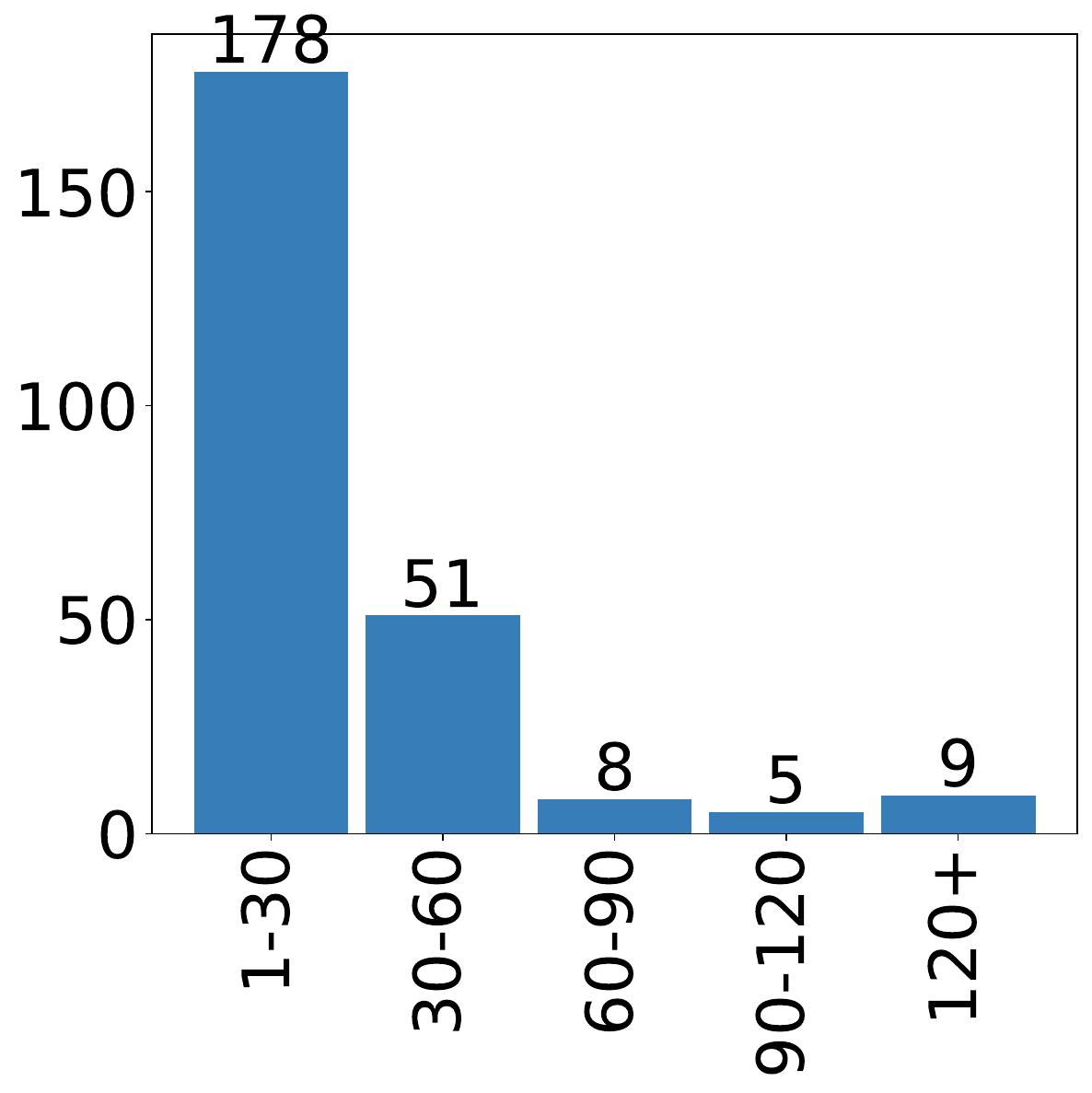}
        \caption{\wcount}
        \label{fig:wcsqlshare}
    \end{subfigure}
    \begin{subfigure}[b]{0.18\textwidth}
        \includegraphics[width=\textwidth]{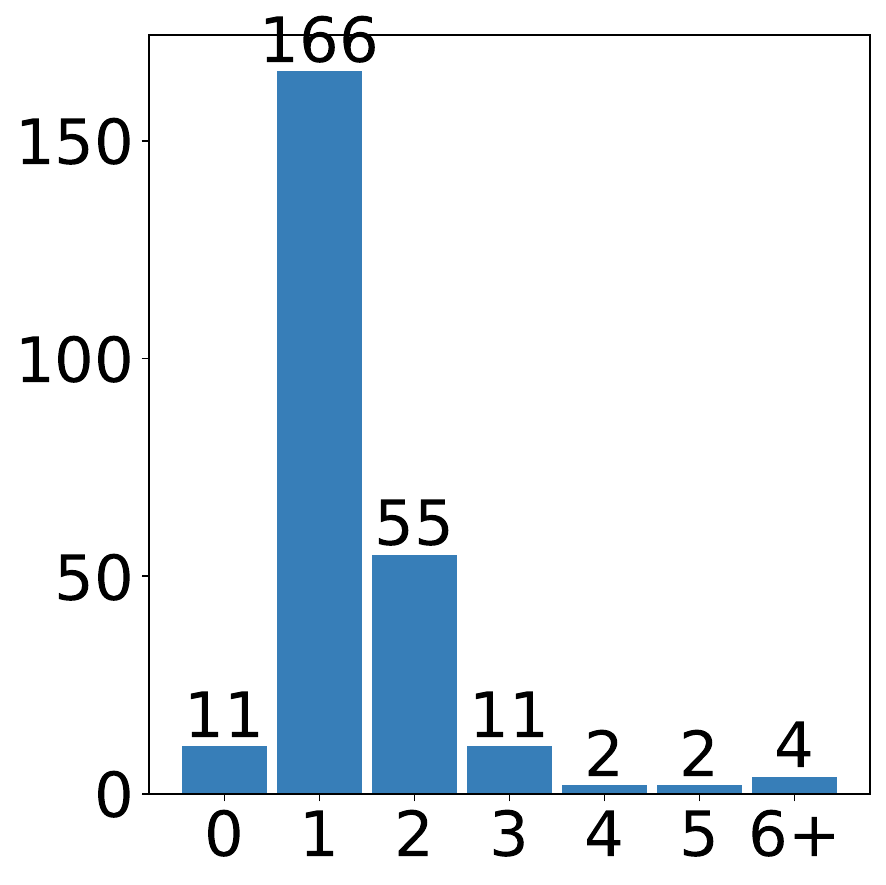}
        \caption{\tcount}
        \label{fig:tcsqlshare}
    \end{subfigure}
    \begin{subfigure}[b]{0.18\textwidth}
        \includegraphics[width=\textwidth]{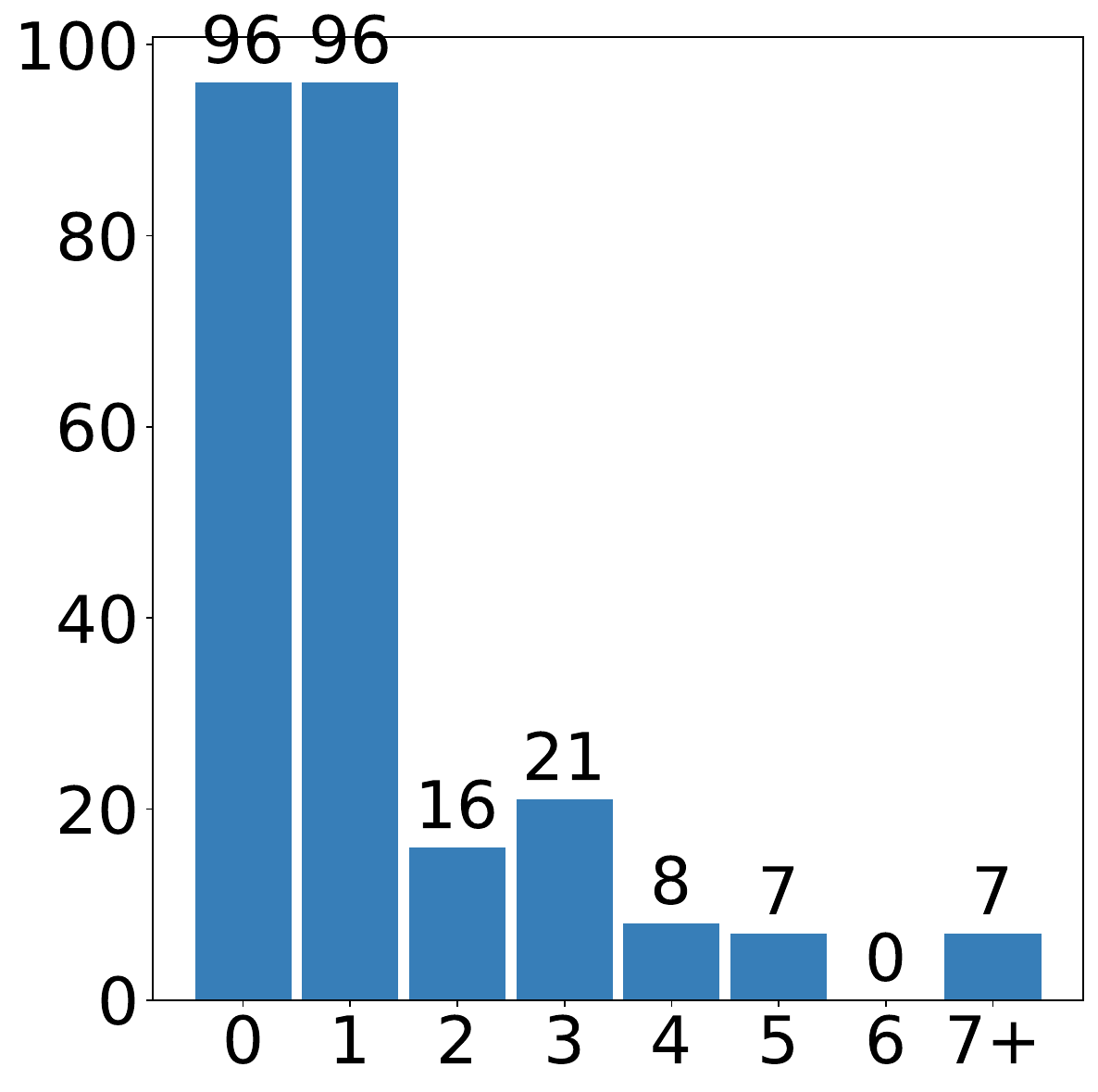}
        \caption{\pcount}
        \label{fig:pcsqlshare}
    \end{subfigure}
    \begin{subfigure}[b]{0.18\textwidth}
        \includegraphics[width=\textwidth]{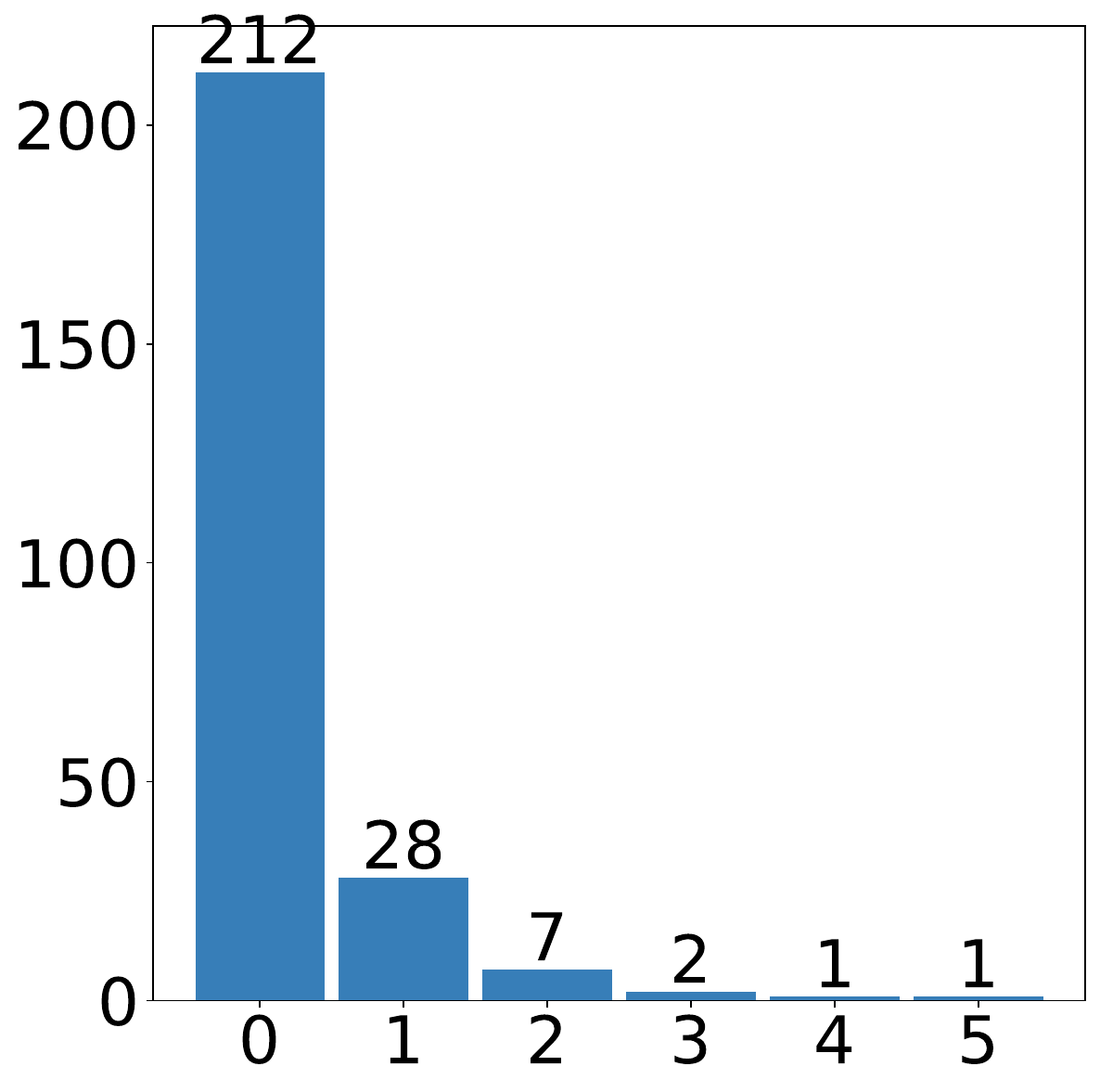}
        \caption{\nlevel}
        \label{fig:nested-sqlshare}
    \end{subfigure}
    \vspace{-3mm}
    \caption{SQLShare Statistics}
    \label{fig:sqlshare}
\end{figure*}

\begin{figure}[h!]
    \centering
    \begin{subfigure}[b]{0.19\textwidth}
        \includegraphics[width=\textwidth]{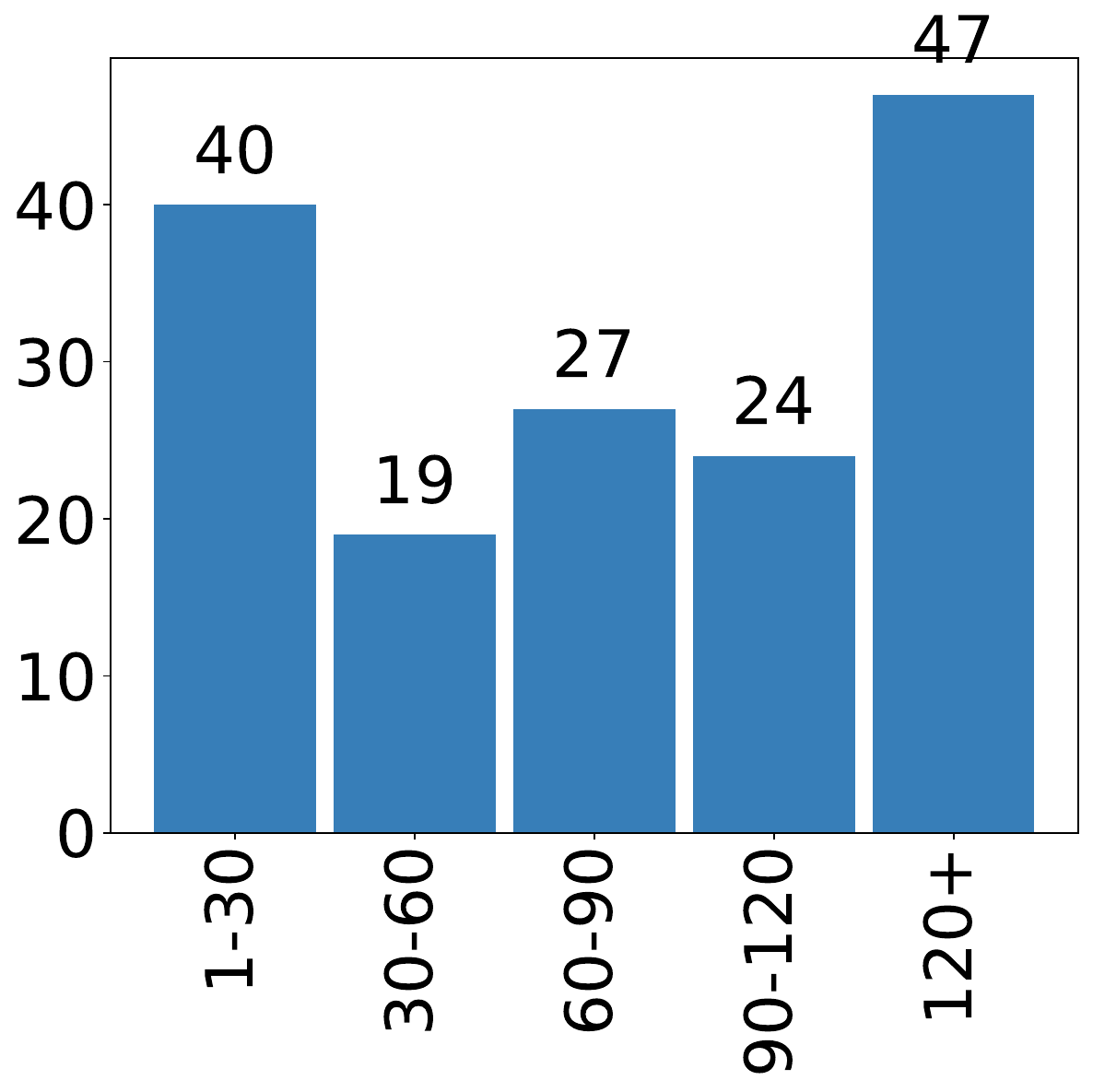}
        \caption{\wcount}
        \label{fig:wcjoin}
    \end{subfigure}
    \begin{subfigure}[b]{0.19\textwidth}
        \includegraphics[width=\textwidth]{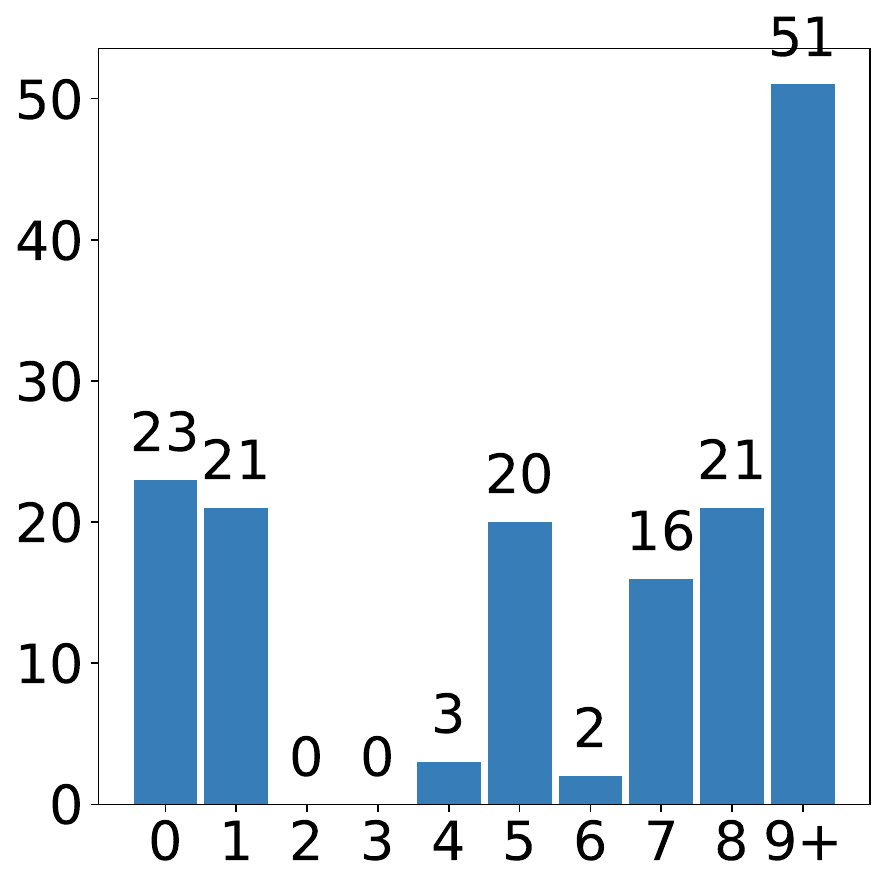}
        \caption{\tcount}
        \label{fig:tcjoin}
    \end{subfigure}
    \begin{subfigure}[b]{0.19\textwidth}
        \includegraphics[width=\textwidth]{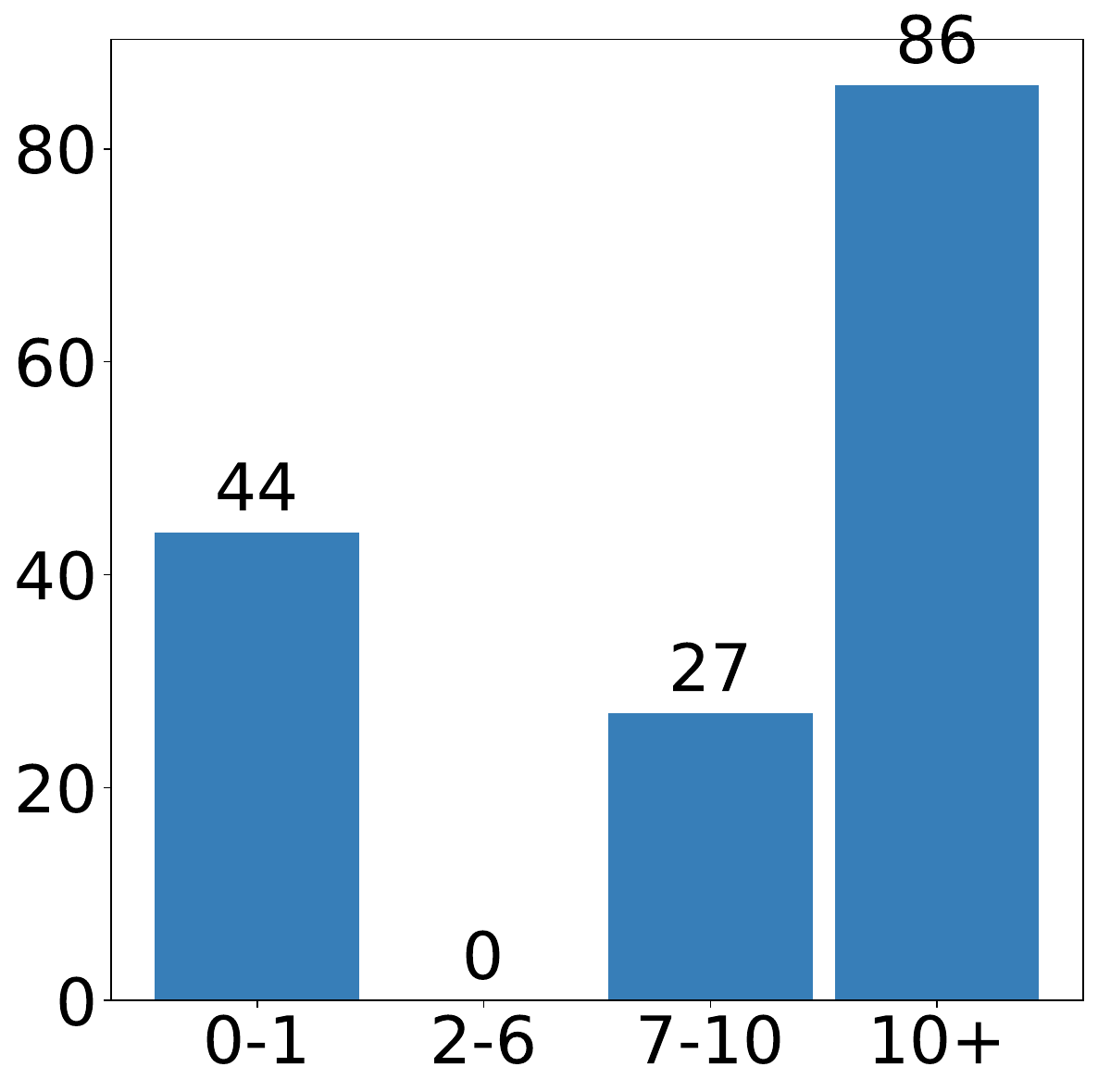}
        \caption{\pcount}
        \label{fig:pcjoin}
    \end{subfigure}
    \begin{subfigure}[b]{0.19\textwidth}
        \includegraphics[width=\textwidth]{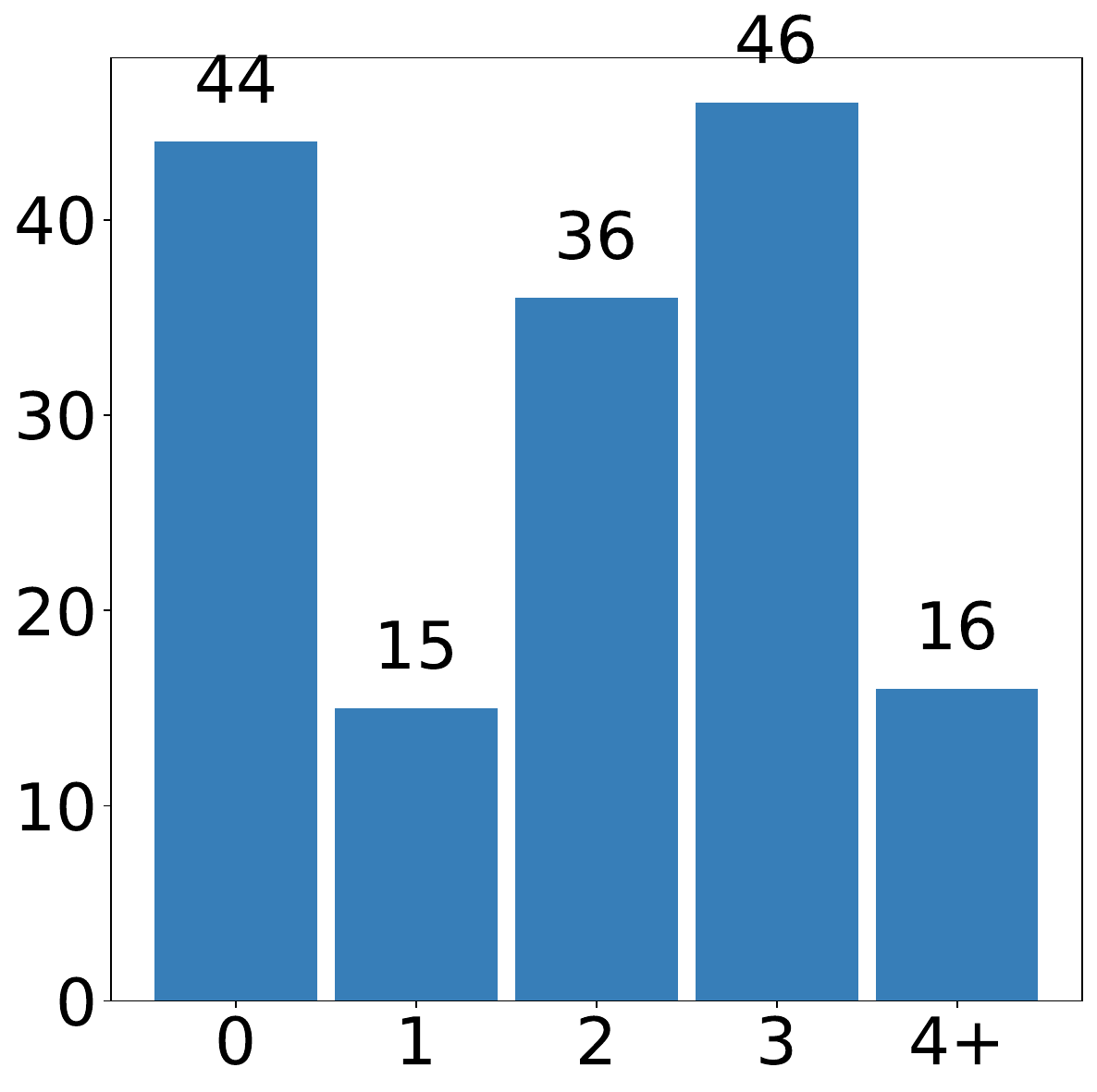}
        \caption{\fcount}
        \label{fig:fcjoin}
    \end{subfigure}
    \vspace{-3mm}
    \caption{Join Order Statistics}
    \label{fig:joinorder}
\end{figure}


\begin{figure*}[h!]
    \centering
    \begin{subfigure}[b]{0.37\textwidth}
        \includegraphics[width=\textwidth]{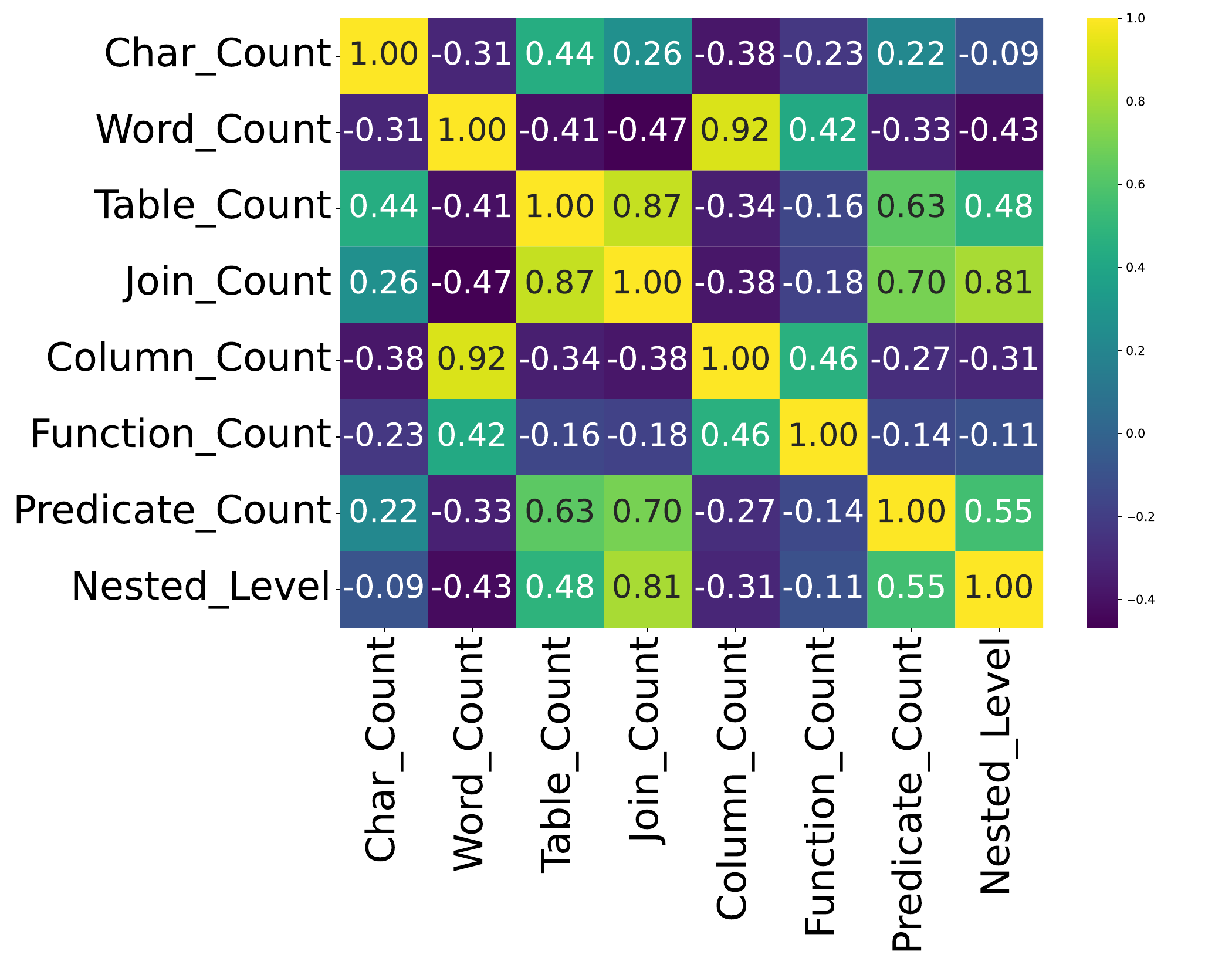}
        \caption{\sdss}
        \label{fig:sdss-stats}
    \end{subfigure}
    \hfill
    \begin{subfigure}[b]{0.29\textwidth}
        \includegraphics[width=\textwidth]{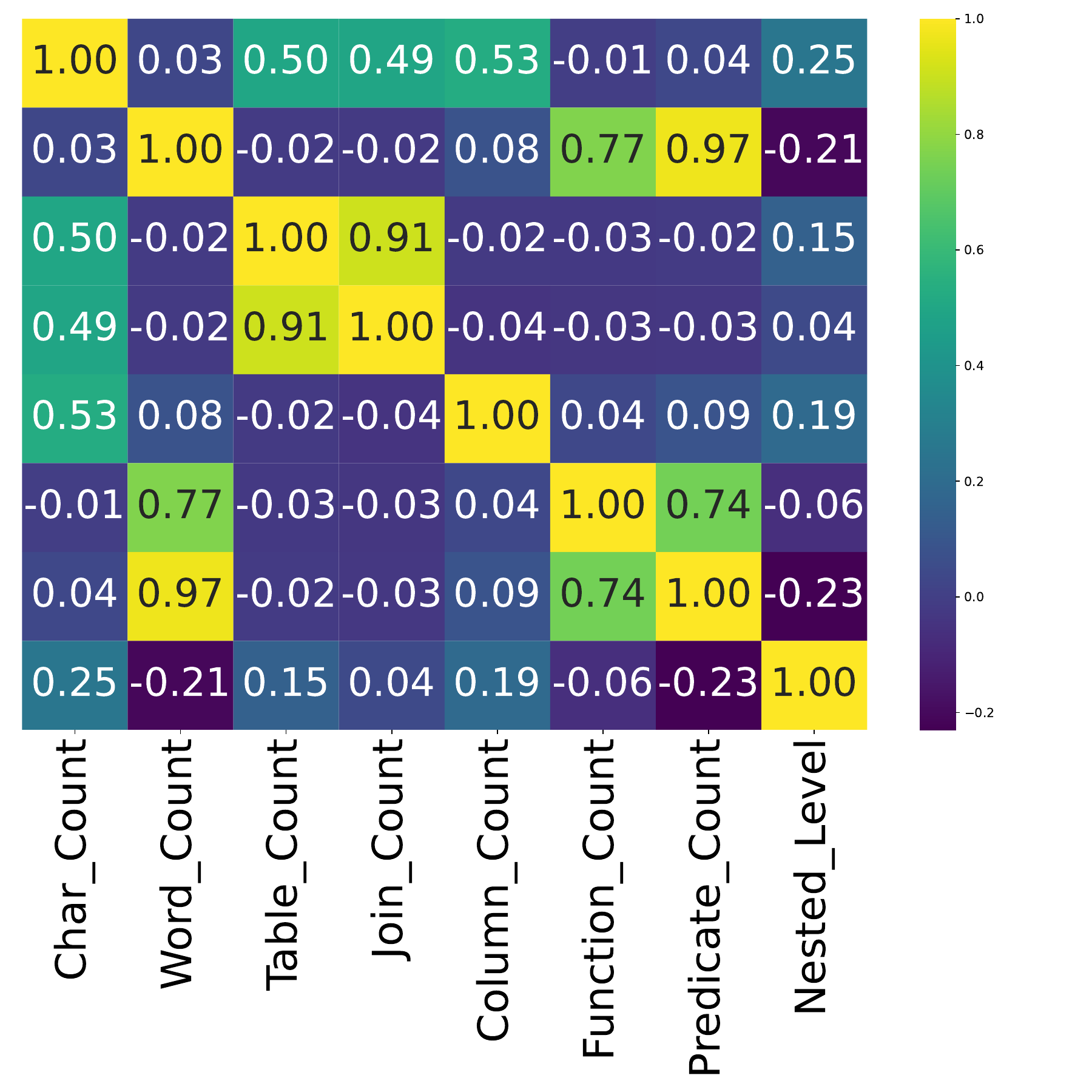}
        \caption{\sqlshare}
        \label{fig:sqlshare-stats}
    \end{subfigure}
    \hfill
    \begin{subfigure}[b]{0.29\textwidth}
        \includegraphics[width=\textwidth]{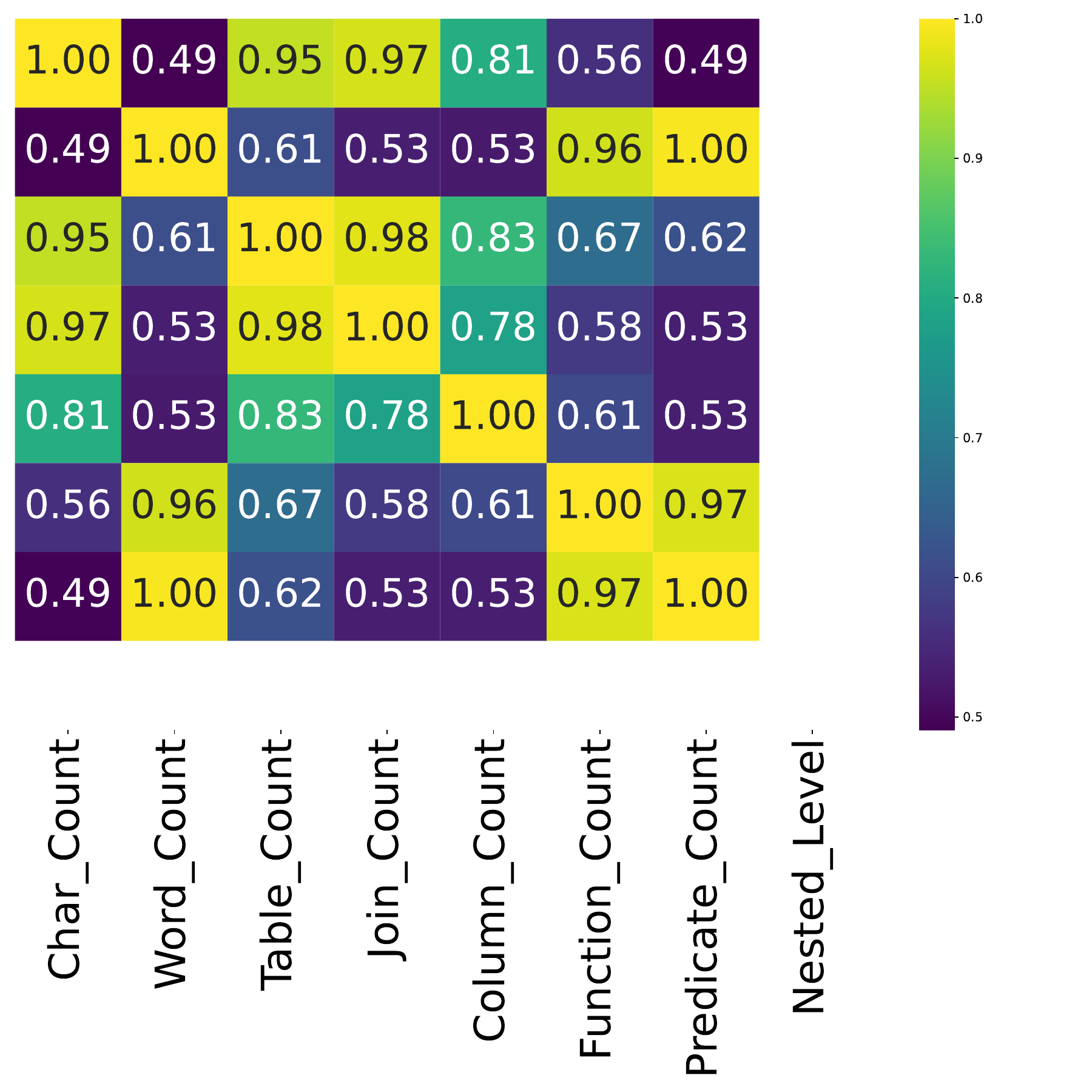}
        \caption{\joinorder}
        \label{fig:join-stats}
    \end{subfigure}
    \vspace{-3mm}
    \caption{Pairwise correlations between query properties for each workload}
    \label{fig:workload-stats}
\end{figure*}

\subsection{Syntactic Properties of SQL Queries}

For each SQL query, we assess the following properties:

\begin{itemize}[nolistsep, leftmargin=*]
    \item \qlen and \wcount, respectively, refer to the number of characters and the number of words in the query.    
    \item \qtype refers to the type of the query, e.g., \texttt{SELECT}, \texttt{UPDATE}, and \texttt{CREATE}.    
    \item \tcount and \jcount refer to the number of distinct tables referenced in the query and the total number of joins, respectively. Joins include both explicit joins (using join keywords such as \texttt{INNER JOIN}) and implicit joins (tables in the \texttt{FROM} clause with join conditions).
    \item \ccount refers the number of distinct columns used or referenced in the SELECT clause of the query.
    \item \fcount refers to the total number of functions in the query, including built-in (like \texttt{min}, \texttt{avg}) and user-defined functions.  \pcount is the number of conditions specified in the \texttt{WHERE} clause.
    \item \nlevel is the nested depth of subqueries within the query.
    \item \qaggregate refers to whether the query uses aggregate functions. 
\end{itemize}



Table~\ref{tab:dataset_stats} provides a statistical overview of all four workloads, including the number of \texttt{SELECT} and \texttt{CREATE} queries and a breakdown of aggregate vs. simple queries. Figures~\ref{fig:sdss}-\ref{fig:joinorder} illustrate additional properties. Each figure is a histogram showing query counts on the $y$-axis and query properties on the $x$-axis, where the $x$-values represent a range of properties. For example, Figure~\ref{fig:wcsdss} shows the number of queries ($y$-axis) across different ranges of query lengths (\wcount). The figures highlight that \sdss and \sqlshare contain more complex queries, with multiple tables and a wider variety of predicates. In contrast, \joinorder has simpler, less nested queries. For query length (\wcount), \sdss and \joinorder have longer queries compared to \sqlshare.

\eat{\textcolor{green}{frequency counts?  We should label the y-axes to be clear for readers who may miss the text}}

As pairs of properties may exhibit strong correlations, leading to redundancy and inefficiency, we examine the correlations between pairwise query properties using Pearson coefficients~\cite{pearson1895}, and we use a threshold of 0.7 to indicate strong correlation.  Figure~\ref{fig:workload-stats} show the following observations:



\begin{itemize}[nolistsep, leftmargin=*]
\item \qlen and \wcount are highly correlated, as longer queries generally contain more words.
\item \tcount and \jcount are also highly correlated since queries with more tables usually involve more joins, a common pattern in multi-table SQL queries.
\end{itemize}

\noindent We consider correlations unique to specific workloads:

\begin{itemize}[nolistsep, leftmargin=*]
\item In \sdss, \ccount and \qlen are strongly correlated as longer queries often involve selecting more columns or adding conditions. Additionally, \nlevel and \jcount are correlated, as deeply nested queries tend to include multiple joins. 
\item \sqlshare and \joinorder exhibit a high correlation between \fcount and \pcount due to the frequent use of functions in conditions. 
\item In \joinorder, \qlen, and \wcount are correlated with \tcount and \jcount, indicating longer queries involve more joins and tables. 
\end{itemize}

\section{Experimental Setup}\label{sec:setup}

We introduce our SQL tasks in Section~\ref{sec:tasks}, and our data preparation steps to inject errors, missing tokens, and derive equivalent and non-equivalent queries in Section~\ref{sec:prepare}.  We then give an overview of the evaluated LLMs (Section~\ref{sec:LLMs}), and how we prompt and respond to the LLMs in Section~\ref{sec:promptEng}.


\subsection{SQL Tasks} \label{sec:tasks}


\subsubsection{Binary Tasks} We begin with binary classification tasks that identify syntactic errors, missing tokens, and query equivalence.  

\paraStart{\synerror.} We evaluate the LLM's ability to identify the presence of a syntax error. We study six types of errors as descrbed below.   Listing~\ref{query:merged} shows sample errors for each type. 

\begin{itemize}[nolistsep, leftmargin=*]
    \item \texttt{aggr-attr.} Aggregate functions are used without properly grouping non-aggregated columns.
    \item \texttt{aggr-having.} Misusing the \texttt{HAVING} clause to filter non-aggregated columns instead of using \texttt{WHERE}.
    \item \texttt{nested-mismatch.} The inner query in a nested query returns multiple rows, which is not correctly handled in the outer query.
    \item \texttt{condition-mismatch.} Operations with incompatible data types, e.g., comparing numeric columns to strings.
    \item \texttt{alias-undefined.} An alias is used in a query but is not defined.
    \item \texttt{alias-ambiguous.} The same column appears in multiple tables, but its usage in a query does not specify the table reference. 
\end{itemize}

\begin{lstlisting}[label={query:merged},caption={SQL syntax error examples}]
-- Q1: Aggregation without GROUP BY (aggr-attr)
SELECT plate,mjd,COUNT(*), AVG(z)
FROM SpecObj WHERE z > 0.5;
-- Q2: Incorrect Use of HAVING (aggr-having)
SELECT plate,COUNT(*) AS NumSpectra
FROM SpecObj GROUP BY plate HAVING z > 0.5;
-- Q3: Type mismatch in subquery (nested-mismatch)
SELECT p.ra,p.dec,s.z
FROM PhotoObj AS p JOIN SpecObj AS s
ON s.bestobjid = (SELECT bestobjid FROM SpecObj);
-- Q4: Type mismatch in condition (condition-mismatch)
SELECT plate,mjd,fiberid FROM SpecObj WHERE z = 'high';
-- Q5: Undefined alias (alias-undefined)
SELECT s.plate,s.mjd,z 
FROM SpecObj AS s JOIN PhotoObj AS p
ON s.bestobjid = photoobj.bestobjid;
-- Q6: Ambiguous alias (alias-ambiguous)
SELECT plate,fid FROM SpecObj AS s JOIN PhotoObj AS p
ON s.bestobjid = p.bestobjid WHERE bestobjid > 1000;
\end{lstlisting}

\paraStart{\mtoken.} We probe the LLM to determine whether a SQL query is missing any tokens.  \eat{This binary task is to decide whether a SQL query is missing any tokens.} Although this can be seen as a type of \synerror, we consider it separately due to its importance in applications. We consider six types of tokens: \texttt{keyword}, \texttt{table}, \texttt{column}, \texttt{value}, \texttt{alias}, and \texttt{predicate} (comparisons).

\paraStart{\qequiv.} Determines whether two SQL queries are equivalent, i.e., whether they have the same schema and produce the same results. We study ten types of equivalences and eight types of non-equivalences. Listings~\ref{query:equiv} shows a few examples of these types using the SDSS workload. For the full list of equivalence and non-equivalence types, along with detailed explanations and examples, we refer the reader to our GitHub repository.


\begin{itemize}[nolistsep, leftmargin=*]
    \item \texttt{swap-subqueries.} Swapping inner and outer sub-queries in nested queries.
    \item \texttt{join-nested.} Converting a join into a subquery or vice versa.
    \item \texttt{cte.} Rewriting a query using common table expressions (CTEs), a temporary result set defined using \texttt{WITH}, which simplifies complex queries and is referenced within the main query.
    \item \texttt{reorder-conditions.} Re-arranging the order of conditions in a \texttt{WHERE} clause.
\end{itemize}

We study four types of non-equivalent transformations:

\begin{itemize}[nolistsep, leftmargin=*]
    \item \texttt{agg-function.} Modifying an aggregate function, e.g., updating to \texttt{SUM} from \texttt{AVG}.
    \item \texttt{change-join-condition.} Modifying the type of join, such as switching from an \texttt{INNER JOIN} to a \texttt{LEFT JOIN}.
    \item \texttt{logical-conditions.} Altering logical operators, such as changing \texttt{AND} to \texttt{OR}.
    \item \texttt{value-change.} Updating a filtering condition, e.g., altering the comparison value.
\end{itemize}

\begin{lstlisting}[label={query:equiv},caption={Examples of SQL equivalence and non-equivalence}]
-- Q7: swap-subqueries (Equivalent)
SELECT s.plate,s.mjd FROM SpecObj AS s WHERE s.plate IN
    (SELECT p.plate FROM PhotoObj AS p WHERE p.ra > 180);
-- Equivalent Query:
SELECT p.plate,p.mjd FROM PhotoObj AS p
WHERE p.ra > 180 AND p.plate IN 
    (SELECT s.plate FROM SpecObj AS s);
-- Q8: join-nested (Equivalent)
SELECT s.fiberid FROM SpecObj AS s JOIN PhotoObj AS p 
ON s.bestobjid = p.objid WHERE p.ra > 180;
-- Equivalent Query:
SELECT fiberid FROM SpecObj WHERE bestobjid IN 
    (SELECT objid FROM PhotoObj WHERE ra > 180);
-- Q9: cte (Equivalent)
SELECT plate,mjd FROM SpecObj WHERE z > 0.5;
-- Equivalent Query:
WITH HighRedshift AS 
    (SELECT plate,mjd FROM SpecObj WHERE z > 0.5)
SELECT plate,mjd FROM HighRedshift;
-- Q10: reorder-conditions (Equivalent)
SELECT * FROM SpecObj WHERE plate = 1000 AND mjd > 55000;
-- Equivalent Query:
SELECT * FROM SpecObj WHERE mjd > 55000 AND plate = 1000;
-- Q11: agg-function (Non-Equivalent)
SELECT plate,AVG(z) FROM SpecObj GROUP BY plate;
-- Non-Equivalent Query:
SELECT plate,SUM(z) FROM SpecObj GROUP BY plate;
-- Q12: change-join-condition (Non-Equivalent)
SELECT s.plate,s.mjd FROM SpecObj AS s
JOIN PhotoObj AS p ON s.bestobjid = p.objid;
-- Non-Equivalent Query:
SELECT s.plate,s.mjd FROM SpecObj AS s
LEFT JOIN PhotoObj AS p ON s.bestobjid = p.objid;
-- Q13: logical-conditions (Non-Equivalent)
SELECT plate,mjd,fiberid
FROM SpecObj WHERE z > 0.5 AND ra > 180;
-- Non-Equivalent Query:
SELECT plate,mjd,fiberid
FROM SpecObj WHERE z > 0.5 OR ra > 180;
-- Q14: value-change (Non-Equivalent)
SELECT plate, mjd, fiberid FROM SpecObj WHERE z > 0.5;
-- Non-Equivalent Query:
SELECT plate,mjd,fiberid FROM SpecObj WHERE z > 5;
\end{lstlisting}

\begin{figure}[h!]
    \centering
    \includegraphics[width=0.65\linewidth]{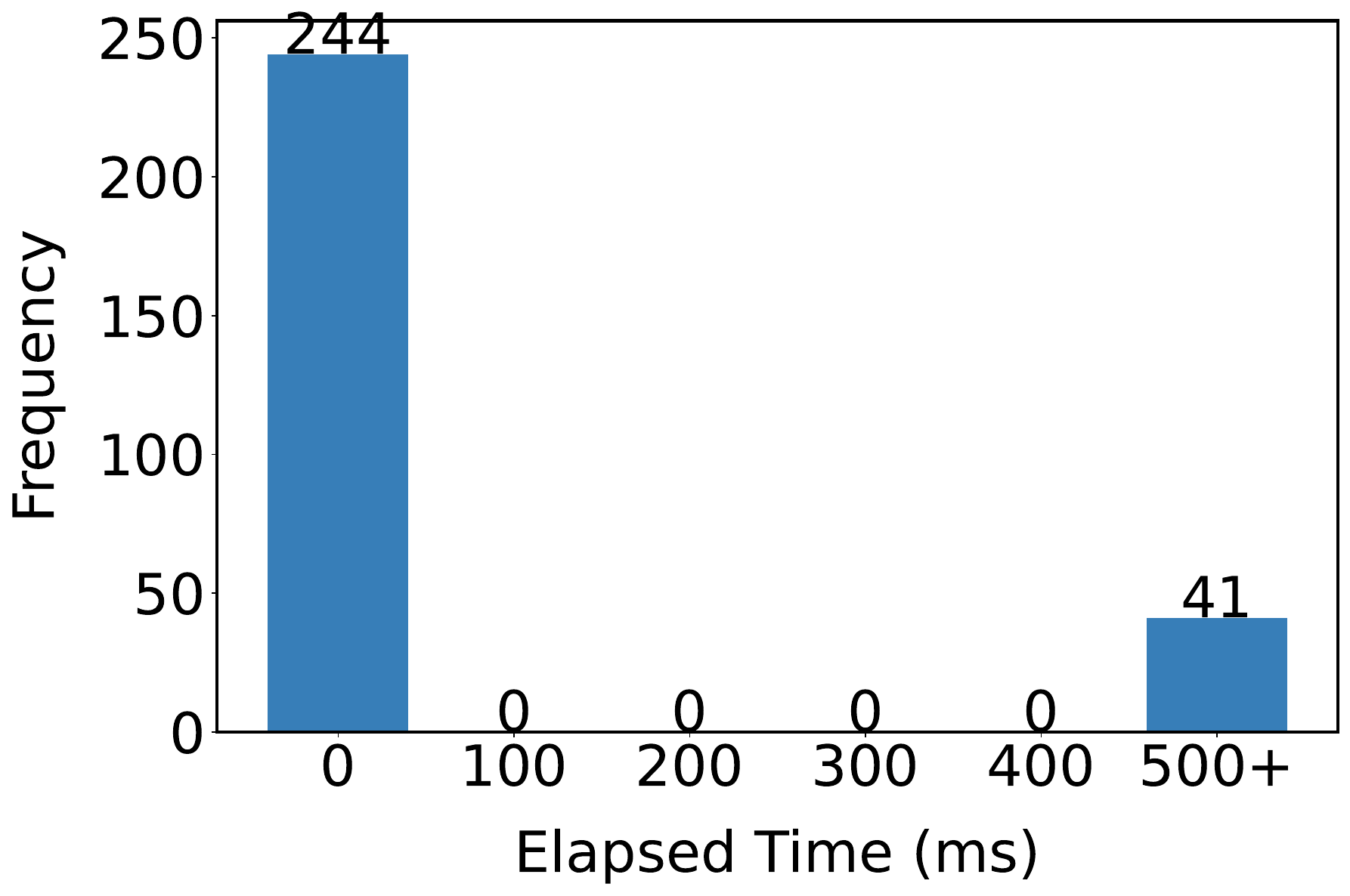}
    \vspace{-3mm}
    \caption{Elapsed time of sampled \sdss queries}
    \label{fig:elapsed}
    \vspace{-2ex}
\end{figure}

\paraStart{\qperformance.} We evaluate the model's ability to predict query runtime performance. Only the \sdss workload contains ground truth query execution times.  Figure~\ref{fig:elapsed} shows a clear separation between short running (low-cost) vs. long running queries (costly), which we pose as a binary classification task, and consider costly queries as the positive class.

\subsubsection{Multi-class Tasks} 
We extend the binary tasks towards multi-class tasks by probing LLMs to indicate the \emph{type} of syntax error (\textbf{\synerror}), type of missing token (\textbf{\mtokent}), and type of query equivalence ({\bf \qequivt}). We also evaluate the task of identifying a missing token's location ({\bf \mtokenl}).

\eat{\feic{I moved performance estimation/prediction to binary tasks. }\mostafa{Looks good.}}


\subsubsection{Query explanation}  This task ({\bf \qexplain}) explains what a SQL query does. It is the reverse of the text-to-SQL task, where existing benchmarks for text-to-SQL, such as WikiSQL~\cite{WikiSQL}, provide natural language descriptions of queries. However, many of these benchmarks contain relatively simple SQL queries compared to the more complex workloads in \sdss and \sqlshare. Thus, we chose the \spider dataset~\cite{Spider} that includes more complex queries, and we further sampled longer and more complex queries.  

Our analysis is qualitative rather than quantitative. We manually review the LLM generated explanations, and compare them with the ground truth descriptions provided in the workload. Our goal is to analyze and discuss when and why models fail to provide accurate, meaningful explanations (Section~\ref{sec:explanation}). While this task does not strictly require existing explanations, we use \spider's explanations to help with validation, and to streamline the evaluation process.

\eat{\ananya {In spider dataset, we have the explanation with the sql queries, we manually check the results of LLMs with the explanation}}

\subsection{Data Preparation and Label Generation} \label{sec:prepare}
We describe how task-specific labels are generated for each dataset.

\vspace{2mm}
\noindent {\bf \synerror} and {\bf \mtoken.} We generate semi-synthetic datasets by randomly selecting queries from each workload and injecting errors. For \synerror, we randomly select the type of error to inject, including cases with no error (error-free).  We create binary labels indicating the presence (or not) of an error, and the error type for the multi-class (\synerrort) task.  We follow a similar method for \mtoken, where we randomly removed a specific token type. We use the \sdss, \sqlshare, and \joinorder workloads for these tasks.



\paraStart{\qequiv.} We use the \sdss, \sqlshare, and \joinorder workloads to generate equivalent and non-equivalent pairs.  We randomly select the type of equivalence, and the queries to modify.  Generating non-equivalent pairs is non-trivial as we must balance sufficient similarity between the queries (to make the task challenging) while including functional differentiation.  Each query pair thus has a label (equivalent/non-equivalent), and a type of equivalence.



\paraStart{\qperformance.} For this task, we used only \sdss, which includes runtime data. We randomly selected 285 queries and analyzed their runtime to classify each as either high or low runtime, serving as a proxy for computational expense. Queries running longer than 200 ms are high cost, otherwise, they are low cost.  We select this threshold based on our observations from Figure~\ref{fig:elapsed}.
 
\subsection{Large Language Models}\label{sec:LLMs}

We use several state-of-the-art LLMs, briefly described below.

\paraStart{\gptthree.} Released by OpenAI in late 2022, \gptthree consists of 175 billion parameters and is trained on a large corpus including Common Crawl, Wikipedia, and various books and academic texts. It is designed to handle diverse NLP tasks~\cite{brown2020gpt3}.

\paraStart{\gptfour.} OpenAI's \gptfour, launched in 2023, boasts over 200 billion parameters, with significant improvements in contextual understanding and reasoning over its predecessor. It was trained on a larger and more diverse dataset, enhancing its performance across a variety of language tasks~\cite{openai2023gpt4}.

\paraStart{\gemini.} Developed by Google and introduced in 2024, \gemini prioritizes both accuracy and safety in AI interactions, with a strong emphasis on ethical AI. It has an estimated parameter count of 50 billion and is trained on vast multimodal datasets, handling both text and visual data. This focus on multimodal capabilities and alignment with human values differentiates it from models primarily designed for text-based tasks~\cite{gemini}.

\paraStart{\llama.} Meta's \llama, released in 2023, is available in versions up to 70 billion parameters. Trained on trillions of tokens from general-purpose datasets, it is designed for efficiency and scalability. LLaMA's focus on broad, general-purpose language tasks makes it ideal for deployment in resource-constrained environments, where maintaining performance \eat{without high computational demands} is crucial~\cite{touvronllama}.

\paraStart{\mistralai.} Launched in 2024, \mistralai is optimized for high accuracy with a smaller footprint of 16 billion parameters. It is trained on a wide range of datasets, with an emphasis on domain-specific content and multilingual capabilities. MistralAI is particularly suited for specialized tasks, that require deeper understanding in areas such as SQL and other structured data, offering a balance of computational efficiency and domain-targeted performance~\cite{mistral2023}.

LLMs achieve high performance due to their extensive training on datasets ranging from hundreds of billions to trillions of tokens. The trend in LLM development is to leverage larger datasets and more complex architectures to continually improve generalization across diverse tasks~\cite{brown2020gpt3,openai2023gpt4,touvronllama,mistral2023}.

\subsection{Refining LLM Interactions} \label{sec:promptEng}
Interacting with LLMs requires careful attention to both input prompts and processing of their responses. By tuning prompts, we guide the models toward generating more accurate and relevant outputs. However, the responses also require post-processing because they are often not provided in a straightforward format necessary for our tasks, such as a simple label. Post-processing involves extracting the necessary information from potentially verbose or complex responses, and ensuring that it fits the specific format required for evaluation.


\paraStart{Prompt Tuning.} Designing and refining input prompts to guide LLMs toward accurate responses is particularly important for complex tasks, where well-crafted prompts can significantly improve model performance~\cite{whiteprompt,sahoo2024systematic,marvin2023prompt}. In our study, prompt tuning is essential to effectively handle the intricacies of SQL syntax and semantics. Our tuning process involved two key steps:

\begin{enumerate}[leftmargin=15pt]
\item {\em Prompt Generation and Refinement.} We used LLMs to generate a variety of prompt candidates, which were then manually refined to ensure clarity and alignment with our task objectives~\cite{wang2024prompt,arawjo2024chainforge}. 
\item {\em Mock Experiments.} We conducted mock experiments with a subset of data to evaluate the effectiveness of each prompt. The top-performing prompts from these tests were selected for full-scale experiments.
\end{enumerate}

Following this approach, we developed a set of task-specific prompts to extract meaningful responses from the models. These prompts were tailored to each experimental task and varied in complexity, addressing challenges such as syntax error detection, query equivalence, and runtime estimation:

\begin{itemize}[nolistsep, leftmargin=*]
    \item \synerror and \synerrort. Does the following query contain any syntax errors? If so, explain the error. [query]
    \item \mtoken, \mtokent, and \mtokenl. Does the following query have any syntax errors? (yes/no) If yes, is there a missing word? (yes/no) If yes, what is the type of the missing word? If yes, what is the missing word? If yes, what is the position of the missing word? (Provide the word count position where the word is missing.) [query]
    \item \qequiv and \qequivt. Are the following two queries equivalent (do they produce the same results on the same database schema)? If yes, why are they equivalent? [query 1, query 2]
    \item \qperformance. Does the following query take longer than usual to run? [query]
    \item \qexplain. Provide a single statement describing this query: [query].
\end{itemize}

The prompts listed above reflect the outcomes of our prompt tuning approach, which was specifically designed to address the SQL tasks in our study.


\paraStart{Handling LLM Output.} Despite providing clear instructions, it is necessary to post-process the LLM output results.  LLMs often \eat{is necessary, even after providing clear instructions, as some LLMs still} produce lengthy and verbose responses that require careful extraction of relevant information. For example, in the prediction task described above, while most LLMs respond with a binary ``yes'' or ``no,'' they often provide explanations about why the query takes a long or short time to execute. Similarly, in the \mtoken task, the responses are not always formatted in a structure that aligns with our evaluation criteria. To address this, we rely on a combination of manual processing and automated scripts. Manual processing involves reading through the responses to extract the specific information required, e.g., isolating the ``yes'' or ``no'' from the rest of the explanation. To expedite this, we use scripts to detect common response patterns and automatically extract the relevant portions when the responses follow predictable structures. However, for more complex or less structured outputs, manual intervention is still necessary to ensure accuracy. This allows us to format the LLM outputs consistently, and to evaluate their performance effectively.



\paraStart{Zero-Shot, Few-Shot, and Fine-Tuning.} \label{sec:zero-few-tune}
\eat{Three main approaches for using LLMs are zero-shot, few-shot, and fine-tuning. }
Zero-shot learning refers to a model’s ability to perform a task without seeing any specific examples, relying solely on its pre-trained knowledge. This approach is valuable for evaluating the model’s inherent understanding of a domain. In our experiments, we focused exclusively on zero-shot learning to assess a model’s ability to detect syntax errors, evaluate query equivalence, and predict query runtime costs. Our goal was to study the models in their raw form, without introducing additional task-specific information, which reflect real-world scenarios where such data may not always be available.

Few-shot learning provides a model with few examples to improve performance, and fine-tuning further trains a model on task-specific datasets to improve accuracy. Although both approaches can help in situations where a model’s initial performance is lacking, we did not use either method in our study. Our goal was to evaluate LLMs with minimal additional training to reflect their performance in environments where limited labeled data are available.







\section{Experimental Results}\label{sec:experiments}

We present our results and analysis, with each subsection focusing on a primary SQL task, and its related secondary tasks.

Across all experiments, \gptfour consistently outperforms other models, with no clear runner-up in most cases. This dominance may be because of the larger model size, as we outlined in Section~\ref{sec:LLMs}, and possibly the model being trained on a larger corpus of SQL queries. To avoid repetition, this general observation will not be restated in the individual result discussions.

\begin{table}[t!]
    \centering
    \setlength{\tabcolsep}{2pt} 
    \begin{tabular}{lcccccccccc}
    \toprule
    \multirow{3}{*}{\rotatebox[origin=c]{90}{\small Case}} & \multirow{2}{*}{Model} & \multicolumn{3}{c}{\sdss} & \multicolumn{3}{c}{\sqlshare} & \multicolumn{3}{c}{\joinorder}  \\
    \cmidrule(lr){3-5} \cmidrule(lr){6-8} \cmidrule(lr){9-11}
     &  & Prec. & Rec. & F1 & Prec. & Rec. & F1 & Prec. & Rec. & F1 \\
    \midrule
    \multirow{5}{*}{\rotatebox[origin=c]{90}{\small Syntax Error}} 
    & \gptfour & \textbf{0.98} & \textbf{0.95} & \textbf{0.97} & \underline{0.94} & \textbf{0.93} & \textbf{0.93} & \textbf{0.95} & \underline{0.91} & \textbf{0.93}\\
    & \gptthree  & 0.94 & 0.85 & 0.89 & {0.91} & {0.86} &{0.89} & \underline{0.93} &{0.81} & {0.86}\\
    & \llama & \underline{0.95} & {0.76} & {0.84} & {0.92} & 0.81 & 0.86 & \textbf{0.95} & {0.65} & {0.77} \\
    & \mistralai & {0.93} & \underline{0.91} & \underline{0.92} &  {0.92} & \underline{0.91} & \underline{0.92} & 0.85 & \textbf{0.94} & \underline{0.89} \\
    & \gemini & {0.94} & {0.70} & 0.80 & \textbf{0.97} & {0.53} & 0.68 & 0.84 & 0.61 & 0.70 \\
    \midrule
    \multirow{5}{*}{\rotatebox[origin=c]{90}{\small Syn. Error Type}} 
    & \gptfour & \textbf{0.96} & \textbf{0.95} & \textbf{0.95} & \textbf{0.89} & \textbf{0.88} & \textbf{0.88} & \textbf{0.90} & \textbf{0.89} & \textbf{0.89} \\
    & \gptthree & {0.87} & {0.85} & {0.85} & \underline{0.85} & \underline{0.82} & \underline{0.83} & {0.83} & {0.78} & {0.78} \\
    & \llama & {0.83} & 0.79 & {0.79} & 0.79 & 0.76 & 0.76 & {0.78} & {0.67} & {0.64} \\
    & \mistralai & \underline{0.90} & \underline{0.88} & \underline{0.89} & {0.81} & {0.80} & {0.79} & \underline{0.86} & \underline{0.81} & \underline{0.82} \\
    & \gemini & 0.81 & 0.74 & 0.73 & 0.73 & 0.60 & 0.58 & {0.68} & {0.53} & {0.52} \\
    \bottomrule
    \end{tabular}
    \caption{Accuracy in \synerror and \synerrort}
    \label{tab:syntax_error_metrics}
    \vspace{-6ex}
\end{table}

\subsection{Syntax Error Tasks} \label{sec:syntax}

In this section, we present results for the two related tasks of \synerror and \synerrort. 

\paraStart{\synerror.} Table~\ref{tab:syntax_error_metrics} (top) shows the comparative accuracy on the \synerror task. The best-performing model is highlighted in bold, and the second-best is underlined. \gptfour, \gptthree, and \mistralai perform well, while \llama and \gemini struggle. This may be because \llama is trained on general-purpose text, and \gemini focuses more on AI ethics and multimodal tasks, meaning both have less specific knowledge of SQL compared to the other models.

Across all models, recall tends to be lower than precision, suggesting that the models are more conservative in detecting errors, missing some existing syntax errors (lower recall) but making fewer incorrect claims about errors (higher precision). One possible explanation is that these models may have been trained more extensively on correct SQL queries, with less exposure to syntactically incorrect examples. This precision-recall imbalance is particularly pronounced in \llama and \gemini, which exhibit significantly lower recall, resulting in reduced F1 scores as well.
 
An important question is when and why LLMs fail in \synerror. To explore this, we examined two hypotheses: first, that failures are related to the syntactic properties of queries, such as \wcount or \tcount; and second, that they are linked to specific types of syntax errors, such as \texttt{aggr-attr} or \texttt{nested-mismatch}, as discussed in Section~\ref{sec:tasks}. We applied the same analysis to other tasks while testing these hypotheses.

\begin{figure}[h]
    \begin{subfigure}[b]{0.22\textwidth}
        \includegraphics[width=\textwidth]{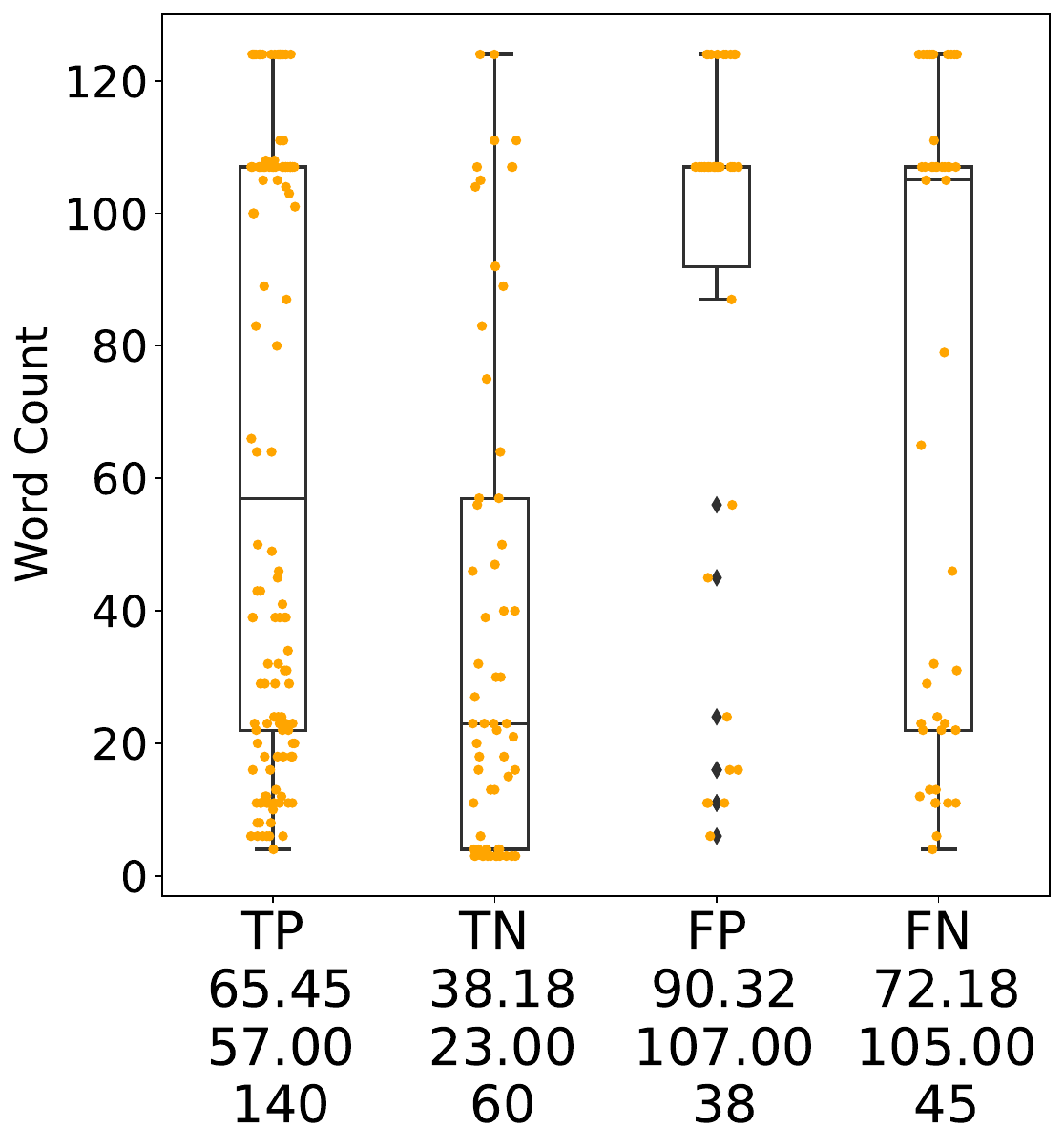}
        \caption{\llama}
        \label{fig:llama-synerror-wcount-sdss}
    \end{subfigure}%
    \begin{subfigure}[b]{0.22\textwidth}
        \includegraphics[width=\textwidth]{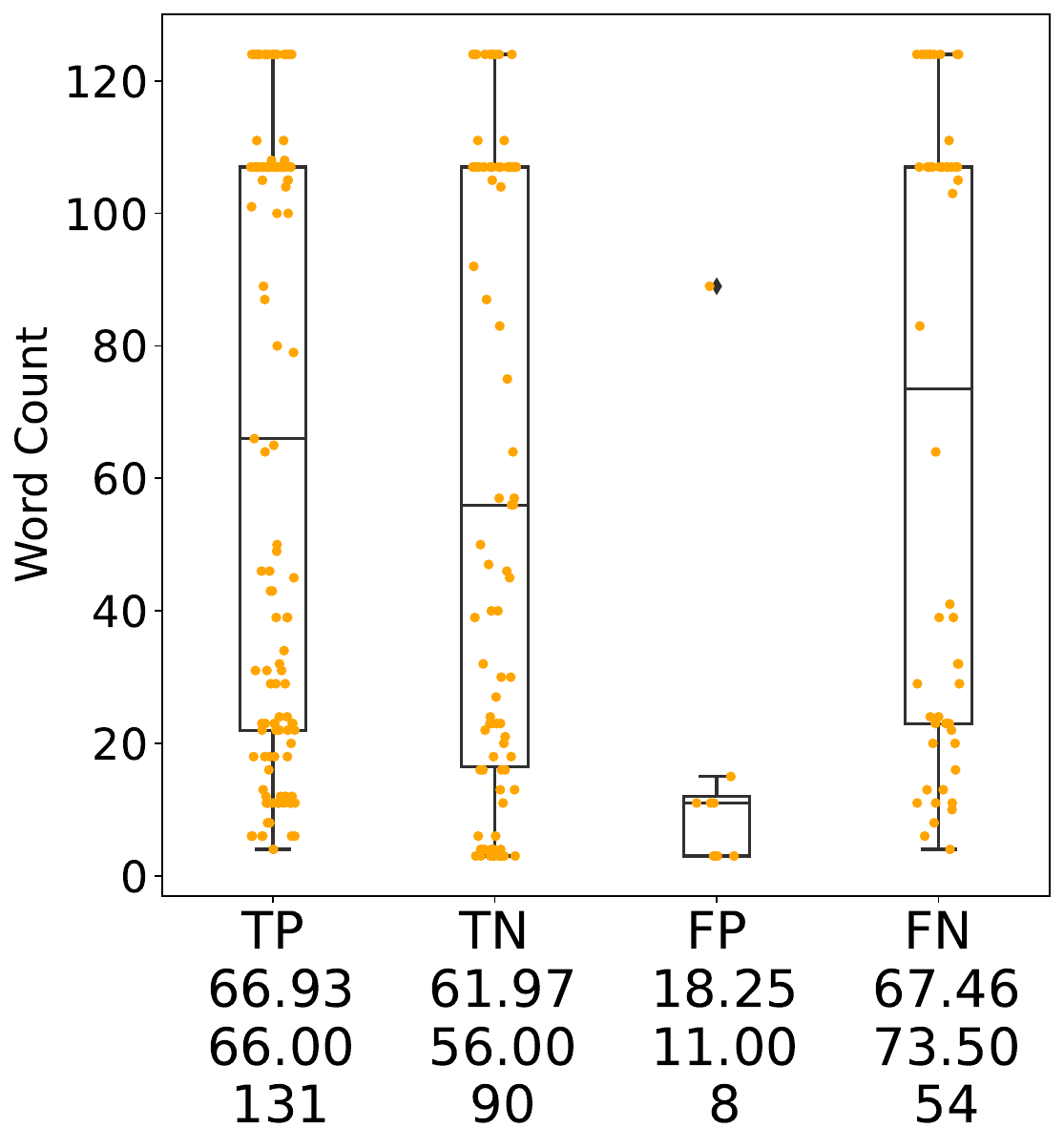}
        \caption{\gemini}
        \label{fig:gemini-synerror-wcount-sdss}
    \end{subfigure}
    \vspace{-3mm}
    \caption{Relationship between \wcount and model failure in \synerror for \sdss. The three numbers (e.g., 65.45, 57.00, 140) represent the average and median query length and the number of queries in the category (\TP). The orange scatter points represent queries plotted on the y-axis by their length, showing the length distribution per category.}
    \label{fig:synerror-wcount-sdss}
\end{figure}

For the first hypothesis, we analyzed the distribution of syntactic properties across four categories: true positive (\TP), true negative (\TN), false positive (\FP), and false negative (\FN). Figure~\ref{fig:synerror-wcount-sdss} illustrates this for \synerror in \sdss, showing query distributions by \wcount. The numbers below each category represent the average (top), median (middle), and total number of queries (bottom). Figures~\ref{fig:llama-synerror-wcount-sdss} and \ref{fig:gemini-synerror-wcount-sdss} show similar data for \llama and \gemini.

To explore the correlation between query length (\wcount) and model failure, we compared \TP and \FN (for queries with errors) as well as \TN and \FP (for queries without errors) while concluding when there are significant queries in each category. For example, in Figure~\ref{fig:llama-synerror-wcount-sdss}, the \TP and \FN categories have sufficient queries (140 and 45, respectively) to observe a pattern: the \FP queries tend to be significantly longer (average 65.45 vs 72.18, median 57 vs 105). A similar trend is seen when comparing \TN and \FP, where \FP queries are longer (average 38.18 vs 90.32, median 23 vs 107). This trend is also observed while comparing \TP and \FN in \gemini in Figure~\ref{fig:gemini-synerror-wcount-sdss}, but there are not enough queries in \FP to draw a conclusion for \TN and \FP. Overall, these findings suggest a correlation between query length (\wcount) and failure in \synerror, with longer queries being more prone to misclassification. We did not observe a similar pattern for any other syntactic properties across models or datasets, indicating that \wcount is the most significant factor influencing failure likelihood in this task.

\begin{figure*}[h]
    \begin{subfigure}[b]{0.32\textwidth}
        \includegraphics[width=\textwidth]{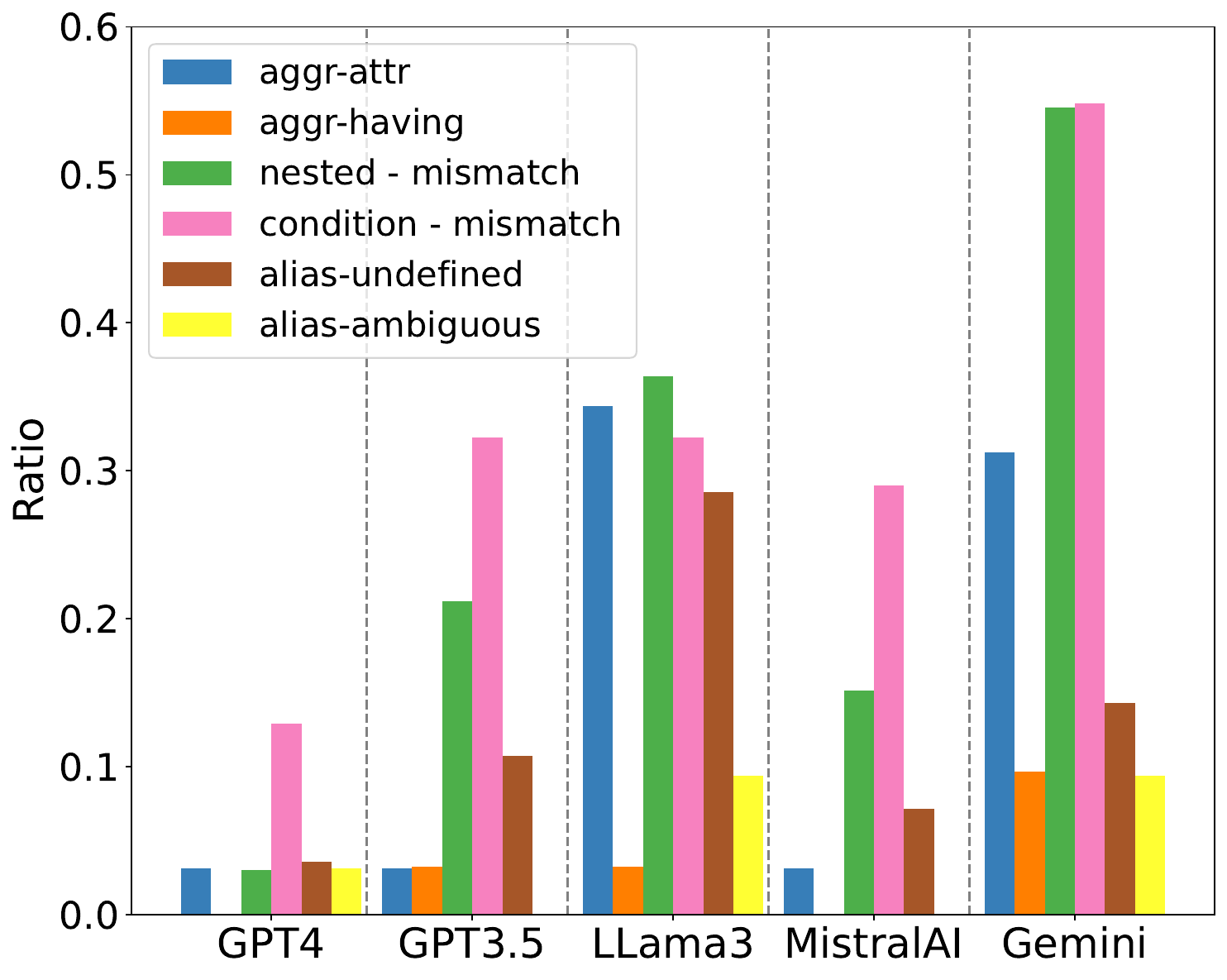}
        \caption{\sdss}
        \label{fig:serror-sdss}
    \end{subfigure}%
    \begin{subfigure}[b]{0.32\textwidth}
        \includegraphics[width=\textwidth]{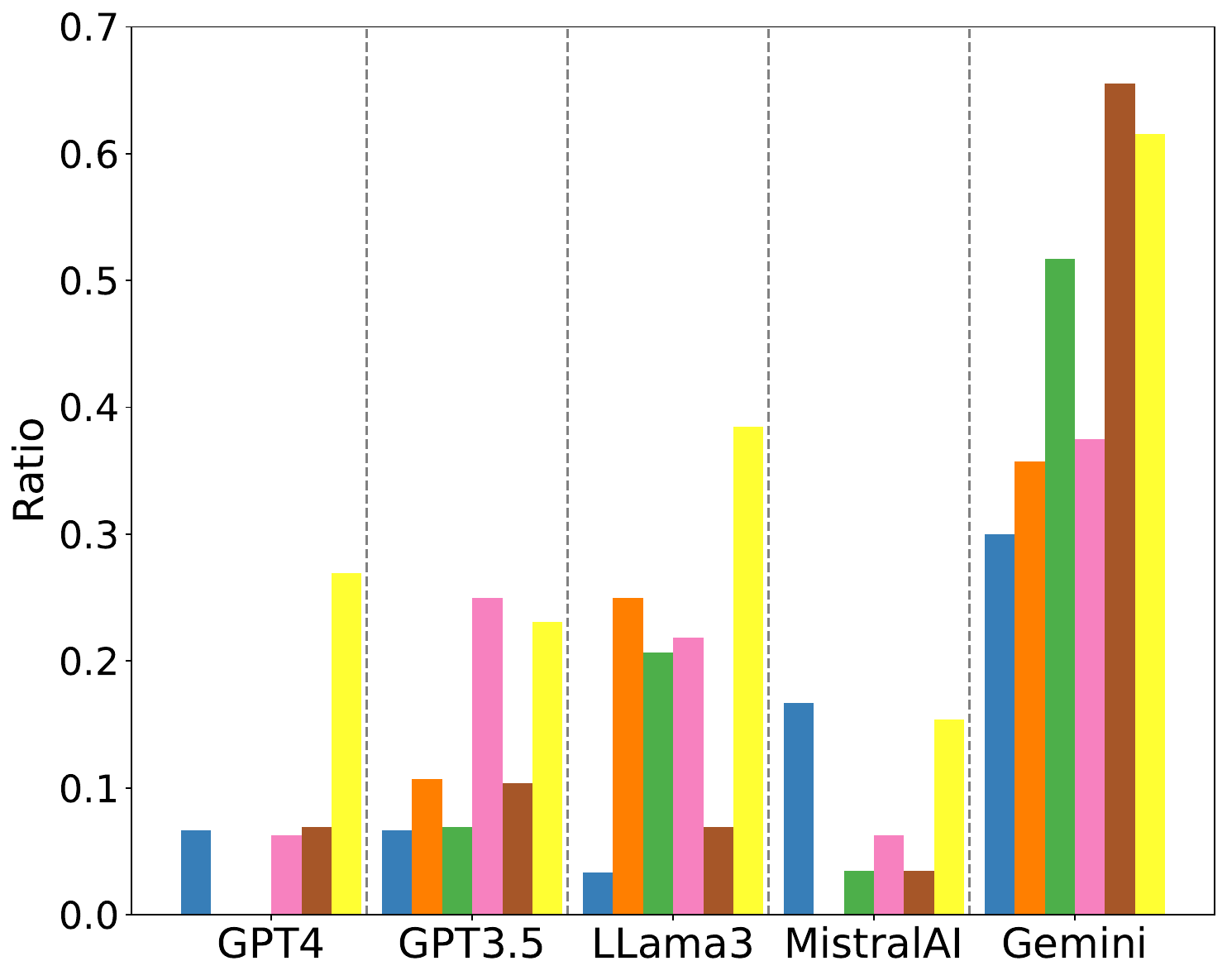}
        \caption{\sqlshare}
        \label{fig:serror-sqlshare}
    \end{subfigure}%
    \begin{subfigure}[b]{0.32\textwidth}
        \includegraphics[width=\textwidth]{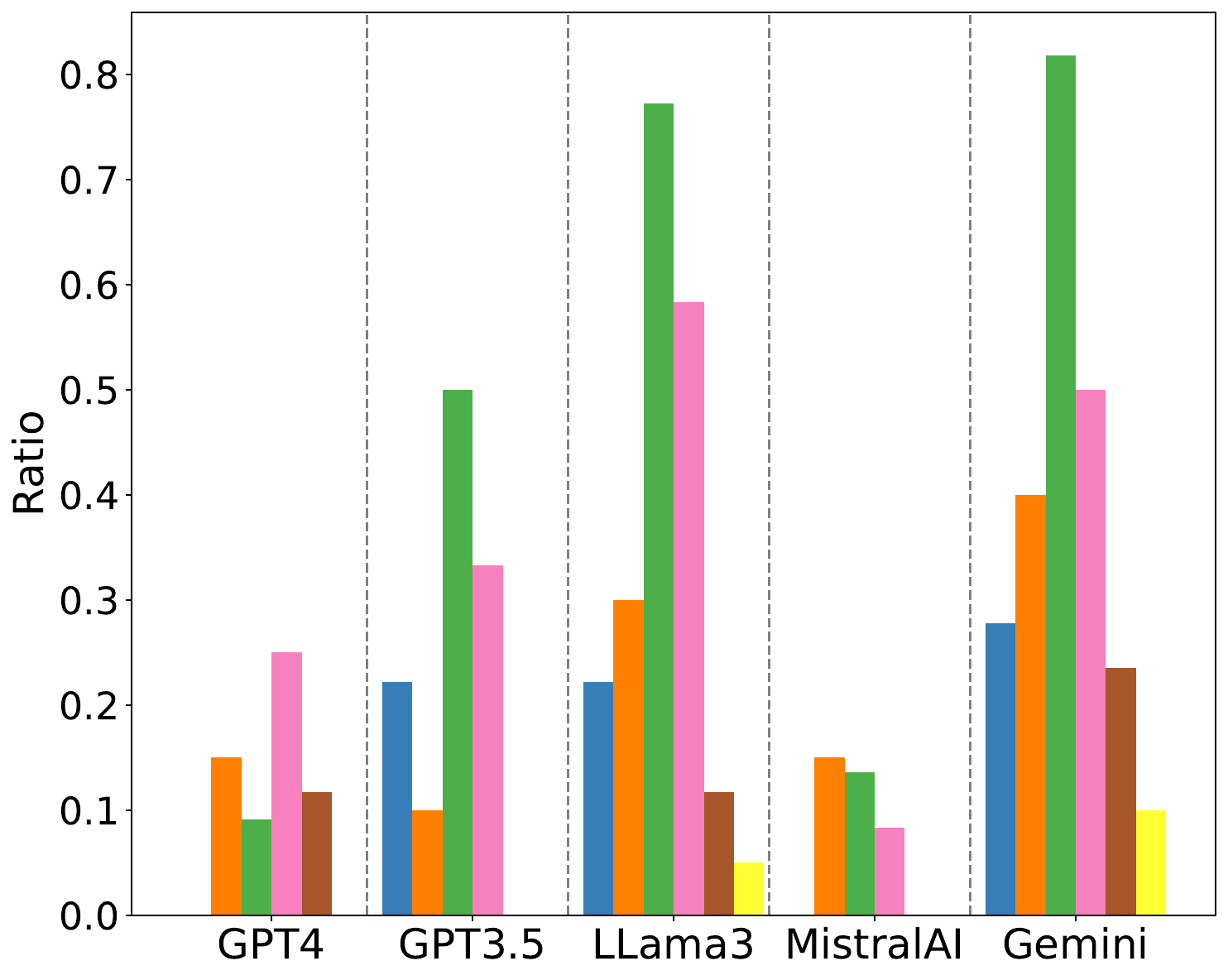}
        \caption{\joinorder}
        \label{fig:serror-join}
    \end{subfigure}
    \vspace{-3mm}
    \caption{Relationship between syntax error type and \FN in \synerror.}
    \label{fig:serror}
\end{figure*}

For the second hypothesis, Figure~\ref{fig:serror} presents the proportion of queries in \FN for each type of syntax error, where a larger bar indicates that detecting errors of that type has been more challenging for the models. The results for \sdss (Figure~\ref{fig:serror-sdss}) suggest that type mismatch errors (\texttt{nested-mismatch} \mybox{mygreen} and \texttt{condition-mismatch} \mybox{mypink}) are particularly difficult for all models to detect. This is expected as the workload involves queries with many conditions for which the type of operands could be difficult to tell. For \sqlshare (Figure~\ref{fig:serror-sqlshare}), ambiguous alias (\texttt{alias-ambigous} \mybox{myyellow}) errors are more problematic, which is expected given the large number of schemas and varied table aliases used in these queries. Lastly, in \joinorder (Figure~\ref{fig:serror-join}), errors involving mismatch in using nested queries (\texttt{nested-mimatch} \mybox{mygreen}) are the most frequently missed by the models since similar to \sdss the queries in \joinorder also have lengthy conditions in the \texttt{WHERE} clauses. 

\paraStart{\synerrort.} Table~\ref{tab:syntax_error_metrics} (bottom) presents the weighted accuracy for \synerrort, which considers the six types of syntax errors (see Section~\ref{sec:tasks}). The strong performance of \gptfour and \mistralai, and \gptthree in detecting syntax errors also extends to identifying error types, while \llama and \gemini continue to perform less effectively, as expected. Overall, results for \synerrort are lower than for \synerror, reflecting the increased difficulty of this task. Another key observation is that all models show lower performance on the \sqlshare dataset, likely due to its more complex schema, which makes identifying the type of syntax errors more challenging.

\takeaway{The analysis of \synerror and \synerrort shows that \gptfour, \mistralai, and \gptthree outperform \llama and \gemini, likely due to differences in training focus. Longer queries are more prone to errors, and the types of syntax errors the models struggle with largely depend on the specific dataset.}

\begin{table}[t!]
    \centering
    \setlength{\tabcolsep}{2pt} 
    \begin{tabular}{lcccccccccc}
    \toprule
    \multirow{3}{*}{\rotatebox[origin=c]{90}{\small Case}} & \multirow{2}{*}{Model} & \multicolumn{3}{c}{\sdss} & \multicolumn{3}{c}{\sqlshare} & \multicolumn{3}{c}{\joinorder} \\
    \cmidrule(lr){3-5} \cmidrule(lr){6-8} \cmidrule(lr){9-11}
     &  & Prec. & Rec. & F1 & Prec. & Rec. & F1 & Prec. & Rec. & F1 \\
    \midrule
    \multirow{5}{*}{\rotatebox[origin=c]{90}{\small Missing Token}} 
    & \gptfour  & \textbf{0.99} & \textbf{0.97} & \textbf{0.98} & \textbf{0.98} & \textbf{0.96} & \textbf{0.97} & \textbf{1.00} & \textbf{0.97} & \textbf{0.99}\\
    & \gptthree  & 0.92 & 0.92 & 0.92 & \underline{0.97} & 0.88 & \underline{0.93} & \underline{0.98} & \underline{0.94} & {0.96}\\
    & \llama & \underline{0.96} & \underline{0.94} & \underline{0.95} & {0.91} & \underline{0.92} & {0.91} & 0.97 & \underline{0.94} & 0.96 \\
    & \mistralai & \textbf{0.99} & {0.86} & 0.92 & {0.96} & {0.87} & 0.91 & \textbf{1.00} & \underline{0.94} & \underline{0.97} \\
    & \gemini & \textbf{0.99} & {0.76} & 0.86 & \textbf{0.98} & {0.68} & 0.80 & 0.97 & {0.69} & 0.81 \\
    \midrule
    \multirow{5}{*}{\rotatebox[origin=c]{90}{\small Token Type}} 
    & \gptfour & \textbf{\accstd{0.94}{0.106}} & \textbf{\accstd{0.94}{0.108}} & \textbf{\accstd{0.94}{0.137}} & \textbf{\accstd{0.91}{0.062}} & \textbf{\accstd{0.89}{0.119}} & \textbf{\accstd{0.90}{0.110}} & \textbf{\accstd{0.98}{0.214}} & \textbf{\accstd{0.97}{0.102}} & \textbf{\accstd{0.98}{0.223}}\\
    & \gptthree  & \accstd{0.76}{0.106} & \accstd{0.75}{0.108} & \accstd{0.75}{0.137} & \accstd{0.75}{0.062} & \accstd{0.71}{0.119} & \accstd{0.73}{0.110} & {\accstd{0.84}{0.214}} & \accstd{0.82}{0.102} & {\accstd{0.82}{0.223}}\\
    
    & \llama & {\accstd{0.88}{0.106}} & \underline{\accstd{0.85}{0.108}} & \underline{\accstd{0.86}{0.137}} & \accstd{0.78}{0.062} & \accstd{0.69}{0.119} & \accstd{0.72}{0.110} & \accstd{0.87}{0.214} & {\accstd{0.82}{0.102}} & \accstd{0.84}{0.223}\\
    & \mistralai & \underline {\accstd{0.89}{0.106}} & \underline{\accstd{0.85}{0.108}} & \underline{\accstd{0.86}{0.137}} & \underline{\accstd{0.82}{0.062}} & \underline{\accstd{0.75}{0.119}} & \underline{\accstd{0.78}{0.110}} & \underline{\accstd{0.93}{0.214}} & \underline{\accstd{0.88}{0.102}} & \underline{\accstd{0.90}{0.223}} \\
    
    & \gemini & \accstd{0.63}{0.106} & \accstd{0.63}{0.108} & \accstd{0.54}{0.137} & \accstd{0.75}{0.062} & \accstd{0.53}{0.119} & \accstd{0.57}{0.110} & \accstd{0.44}{0.214} & \accstd{0.60}{0.102} & \accstd{0.39}{0.223}\\
    \bottomrule
    \end{tabular}
    \caption{Accuracy for \mtoken and \mtokent}
    \label{tab:missing_token_type_metrics}
     \vspace{-6ex}
\end{table}

\subsection{Missing Token Tasks}\label{sec:missing-token-exp}

Regarding missing token, we start by \mtoken, and then present results related to \mtokent and \mtokenl.

\paraStart{\mtoken.} Table~\ref{tab:missing_token_type_metrics} (top) presents the accuracy of various LLMs in \mtoken. Accuracy is higher compared to \synerror, as \mtoken is a simpler task. A notable change is \llama’s improved performance in this task. This can be attributed to the fact that detecting missing tokens relies more on general pattern recognition, which is less specialized for SQL. \llama’s broader training in recognizing patterns likely helps it improve in this context. Overall, recall remains lower than precision in \mtoken, similar to \synerror, likely because the models are more conservative in detecting errors, as explained previously.

\begin{figure}[h]
    \begin{subfigure}[b]{0.22\textwidth}
        \includegraphics[width=\textwidth]{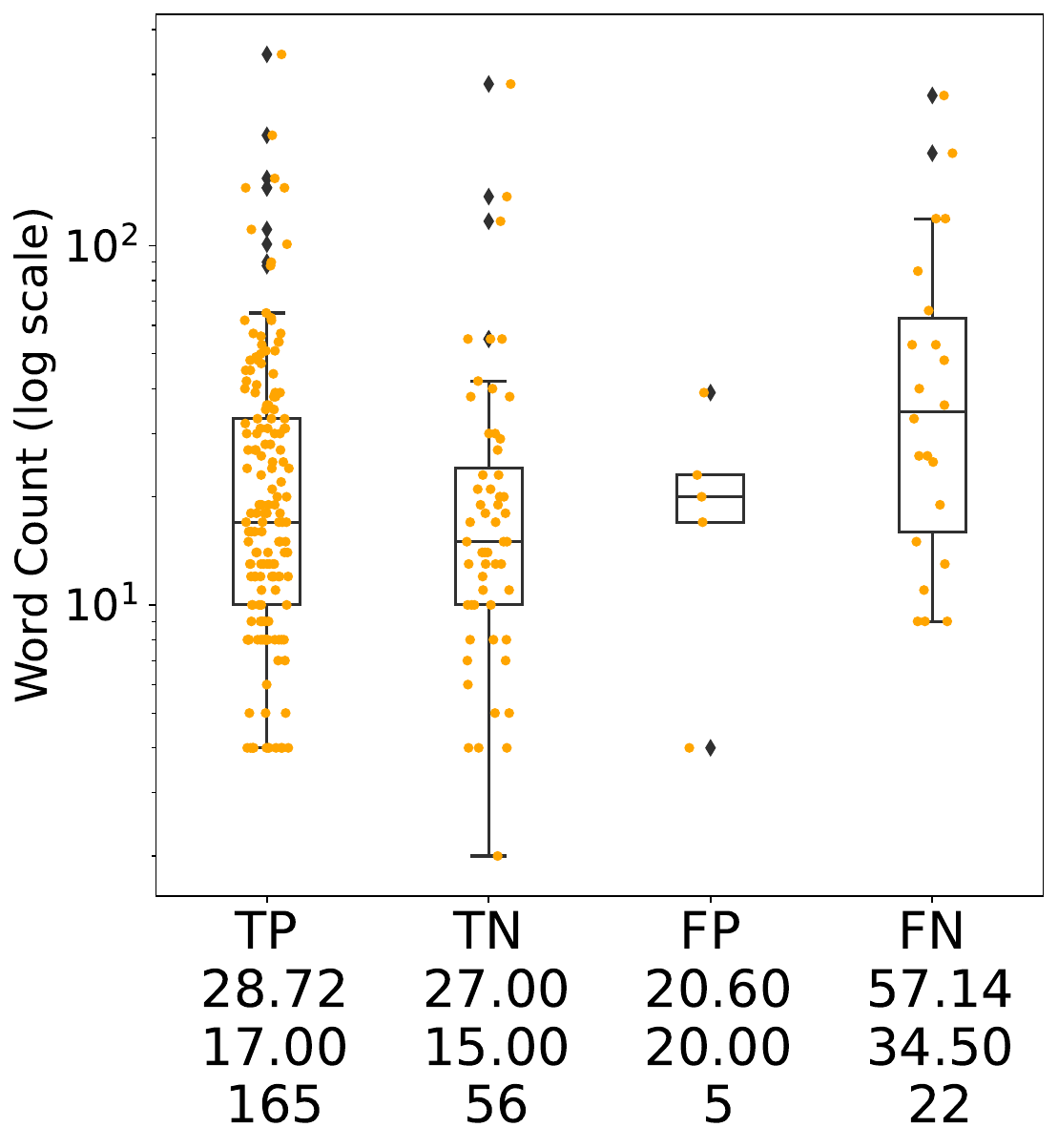}
        \caption{\wcount in \gptthree}
        \label{fig:mtoken-gpt3-wcount}
    \end{subfigure}%
    \begin{subfigure}[b]{0.22\textwidth}
        \includegraphics[width=\textwidth]{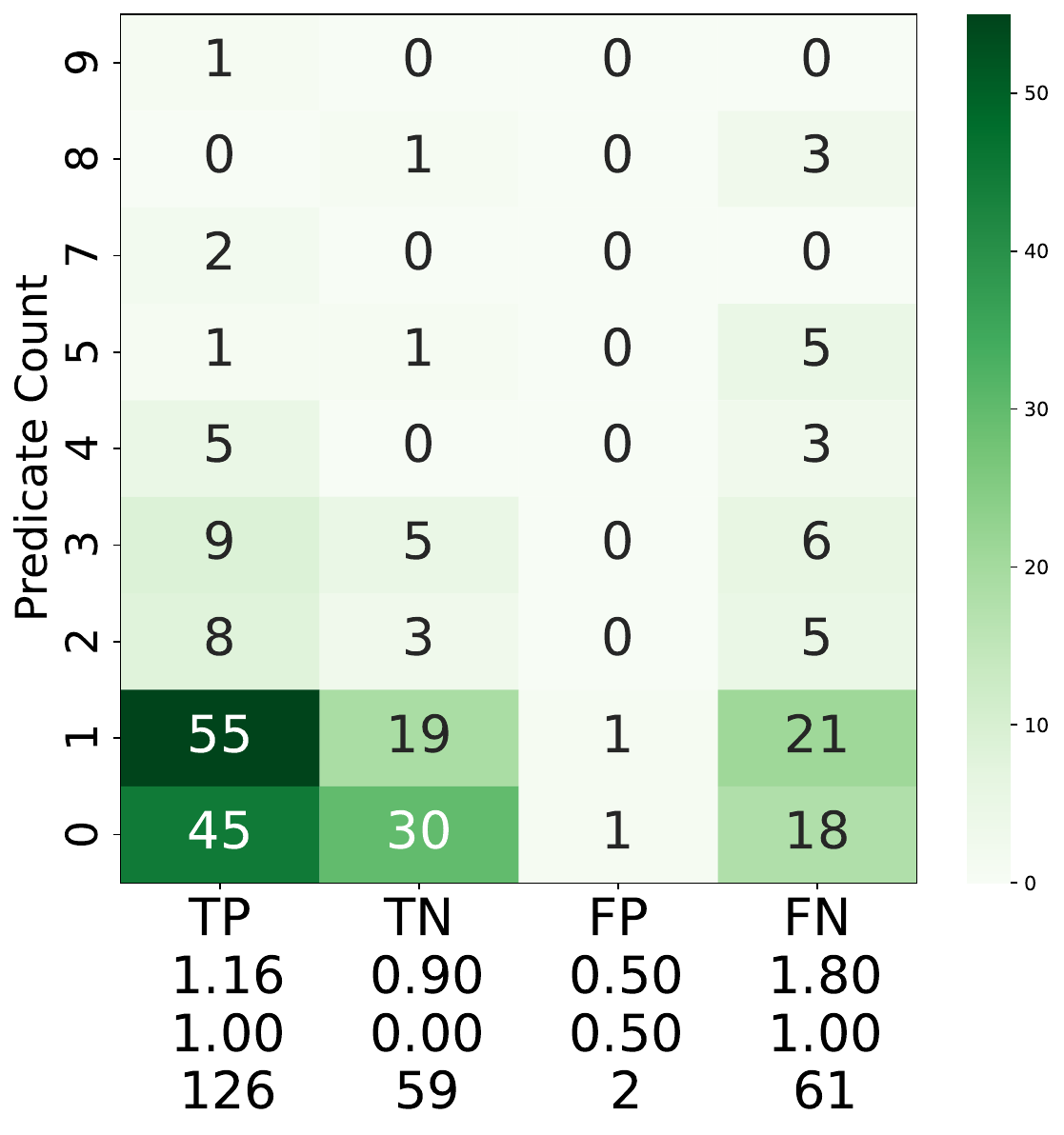}
        \caption{\pcount in \gemini}
        \label{fig:mtoken-gemini-pcount}
    \end{subfigure}

    \begin{subfigure}[b]{0.22\textwidth}
        \includegraphics[width=\textwidth]{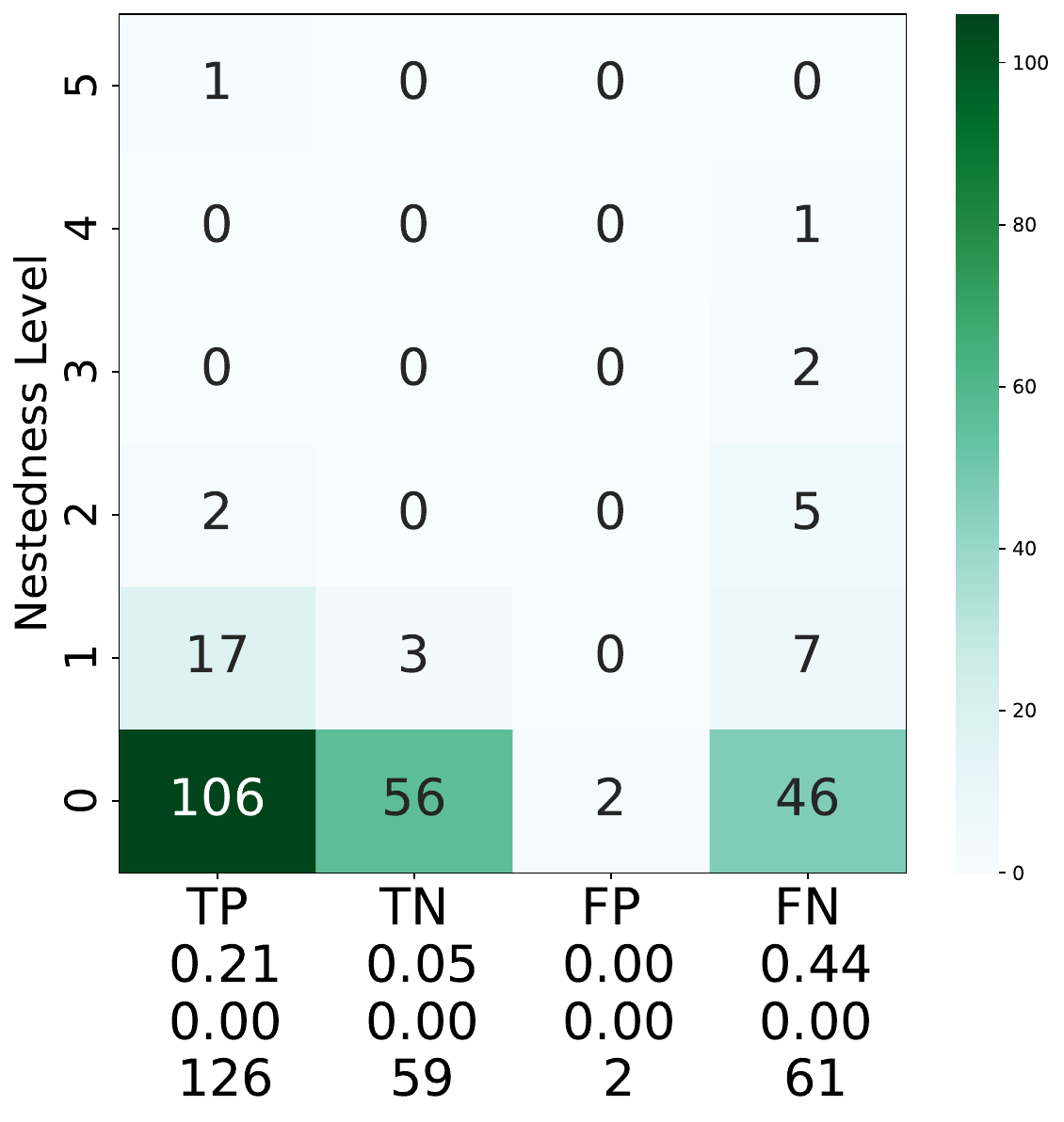}
        \caption{\nlevel in \gemini}
    \label{fig:mtoken-gemini-nlelve}
    \end{subfigure}%
    \begin{subfigure}[b]{0.22\textwidth}
    \includegraphics[width=\textwidth]{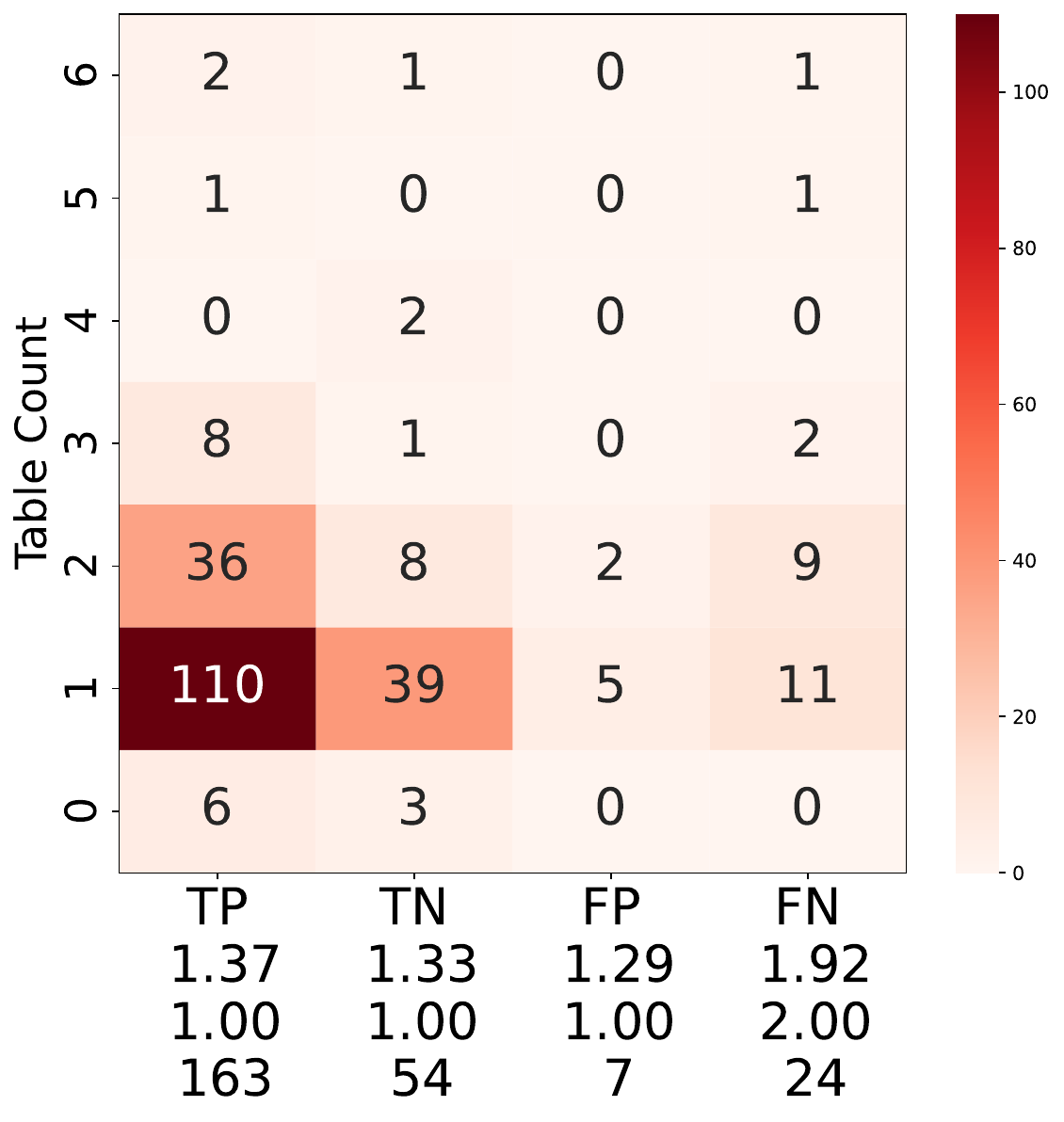}
    \caption{\tcount in \mistralai}
    \label{fig:mtoken-mistralai-table}
    \end{subfigure}
    \vspace{-3mm}
    \caption{LLMs' failure in \mtoken for \sqlshare.}
    \label{fig:correlation-mtoken}
\end{figure}

We investigated the relationship between LLM failures in the \mtoken task and the syntactic properties of queries. Figure~\ref{fig:mtoken-gpt3-wcount} shows that, for \gptthree on the \sqlshare dataset, query length (\wcount) is correlated with failures, with an average \wcount of 57 in \FN compared to 27 in \TP. We also examined other properties such as predicate count (\pcount), nestedness level (\nlevel), and table count (\tcount), as seen in Figures~\ref{fig:mtoken-gemini-pcount},~\ref{fig:mtoken-gemini-nlelve}, and \ref{fig:mtoken-mistralai-table}. In all cases, the average values for \FN are significantly higher than for \TP (1.80 vs 0.90 in \ref{fig:mtoken-gemini-pcount}, 0.44 vs 0.05 in \ref{fig:mtoken-gemini-nlelve}, and 1.92 vs 1.33 in \ref{fig:mtoken-mistralai-table}). However, due to the small number of \FP queries, no definitive conclusions can be drawn for that category.

\begin{figure*}[h]
    \begin{subfigure}[b]{0.32\textwidth}
        \includegraphics[width=\textwidth]{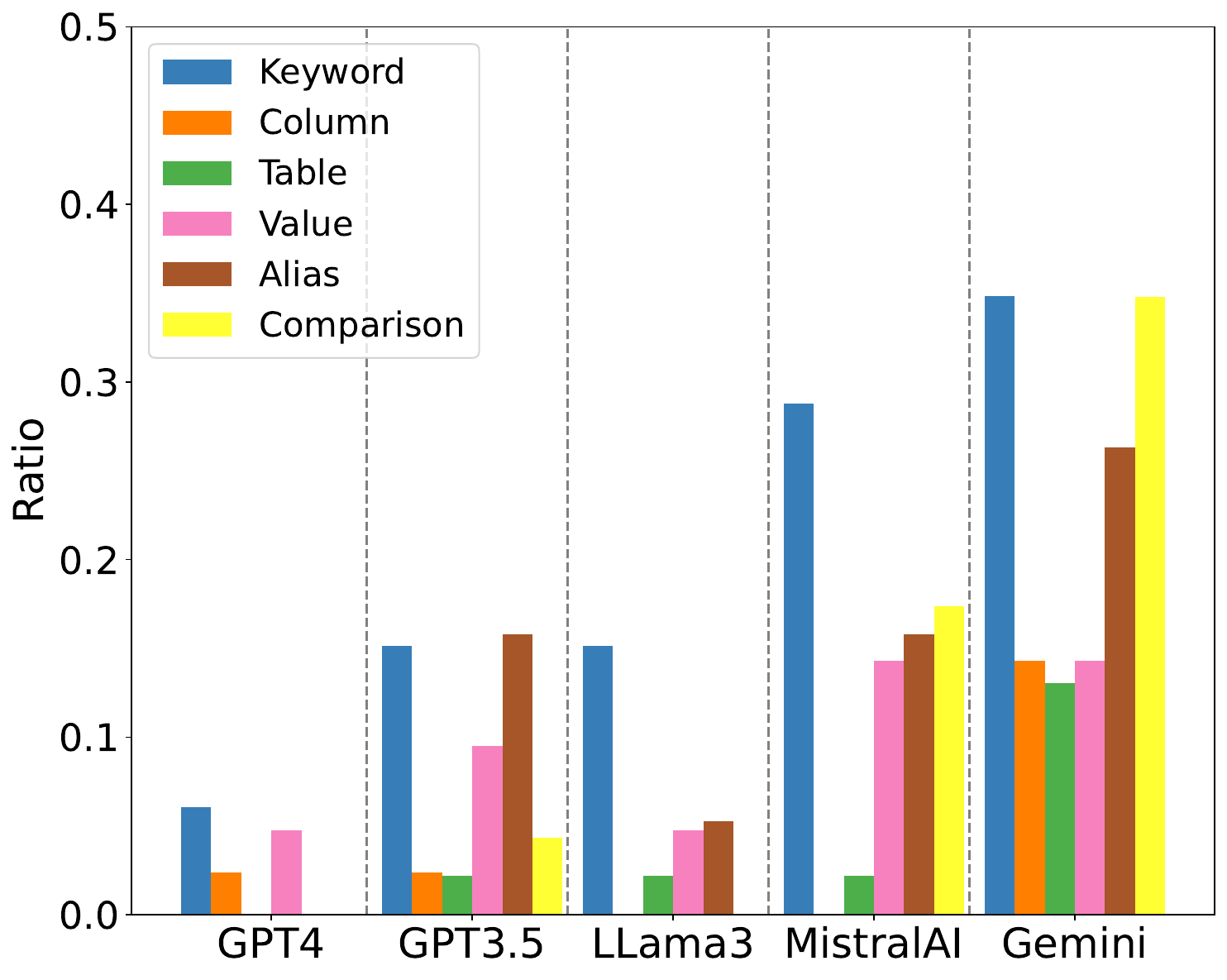}
        \caption{\sdss}
        \label{fig:type-mtoken-sdss}
    \end{subfigure}%
    \begin{subfigure}[b]{0.32\textwidth}
        \includegraphics[width=\textwidth]{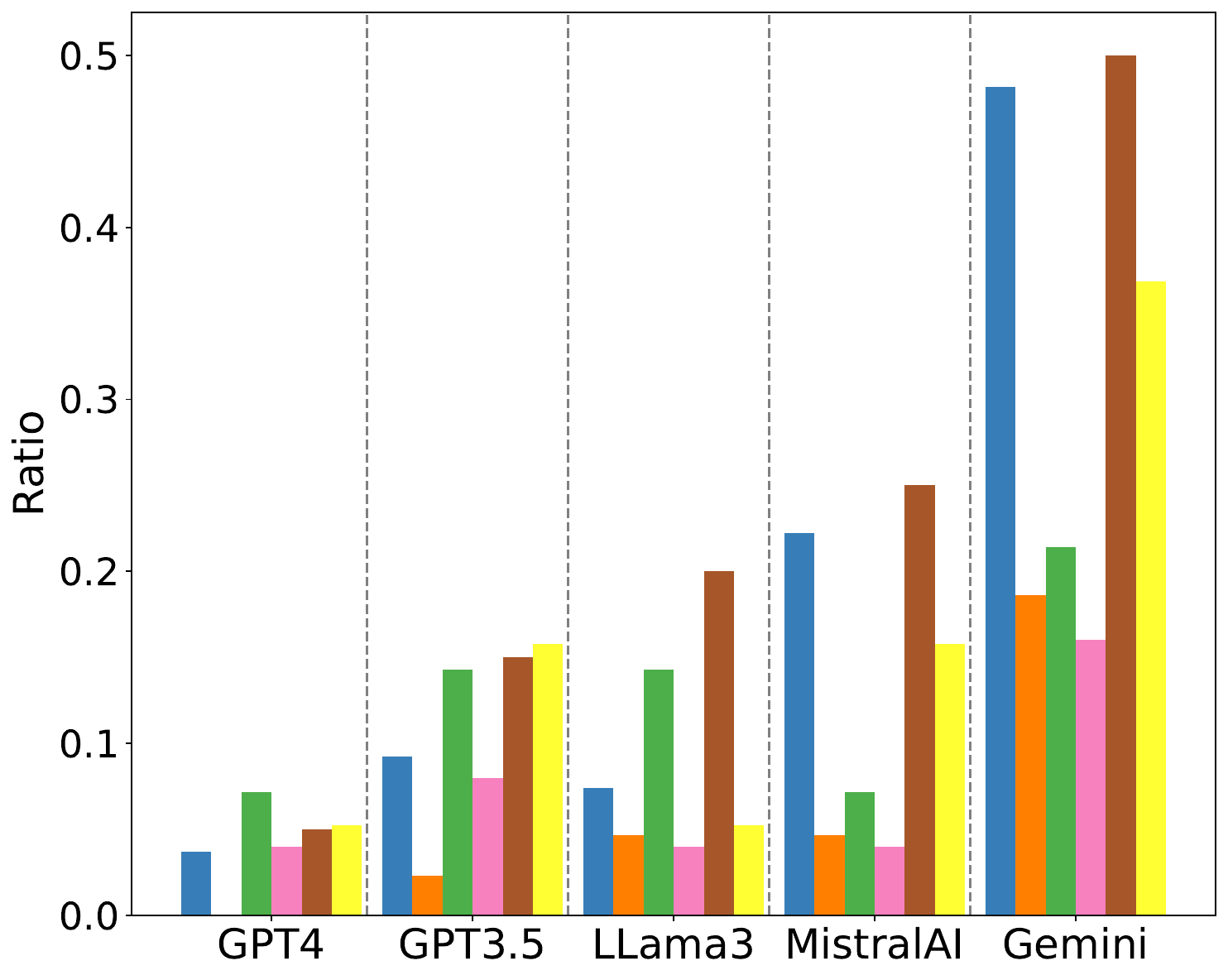}
        \caption{\sqlshare}
        \label{fig:type-mtoken-sqlshare}
    \end{subfigure}%
    \begin{subfigure}[b]{0.32\textwidth}
        \includegraphics[width=\textwidth]{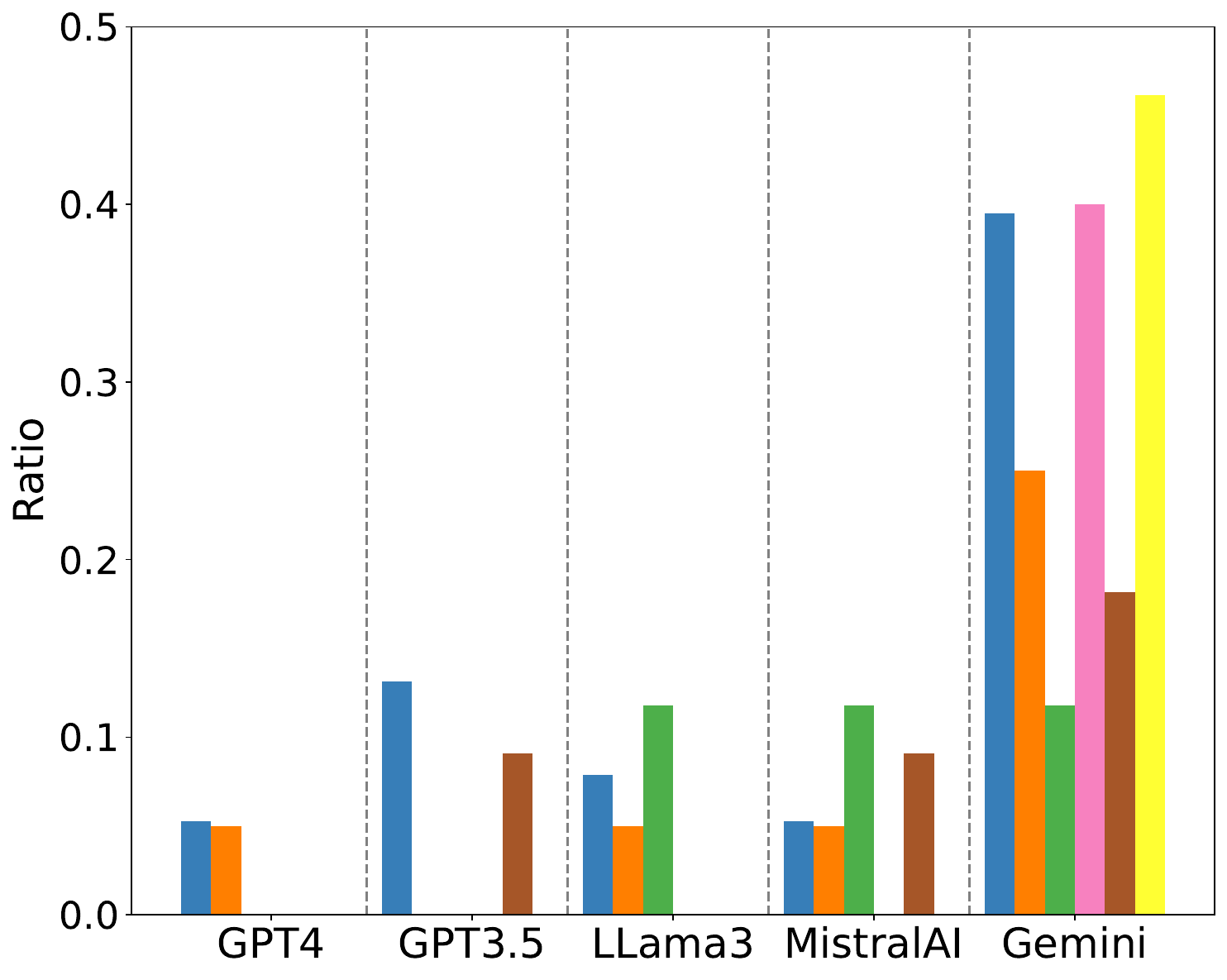}
        \caption{\joinorder}
        \label{fig:type-mtoken-join}
    \end{subfigure}
    \caption{Relationship between missing token type and \FN in \mtoken.}
    \label{fig:type-mtoken}
\end{figure*}

We now shift our analysis to the impact of the missing token type on the performance of LLMs in \mtoken. We examined the breakdown of \FN by token type, as shown in Figure~\ref{fig:type-mtoken}, similar to our analysis for \synerror. A key observation in \sdss is that the most frequent type of failure occurs for \texttt{keyword} (\mybox{myblue}). This is likely because \sdss contains a diverse set of query types with a higher occurrence of keywords compared to \sqlshare and \joinorder. In \sqlshare, the most challenging missing token types are aliases and tables (\mybox{mybrown} and \mybox{mygreen}), which can be attributed to the presence of many small databases with numerous tables and various aliases in their queries. Finally, in \joinorder, there is no single token type with a notably higher failure rate, likely due to the simpler nature of the queries and the relatively low number of failures.

\paraStart{\mtokent.} We reported the weighted average accuracy values in Table~\ref{tab:missing_token_type_metrics} (bottom), with weights based on the number of queries for each type. The results indicate that \mtokent is more challenging than \mtoken across all LLMs, as evidenced by the reduced accuracy. The lowest accuracy is observed in \sqlshare, which is expected due to its complex schema compared to \sdss and \joinorder. Conversely, \joinorder shows the highest accuracy, reflecting its simpler schema. Notably, \mistralai consistently achieves the second-best performance. This is interesting as \llama was the second in \mtoken, which suggests although \llama is better at detection due to its strength in detecting general patterns, \mistralai is better at SQL-related pattern recognition, correctly deciding the error type for more queries.

\begin{table}[t]
    \centering
    \setlength{\tabcolsep}{2pt} 
    \begin{tabular}{cccccccc}
    \toprule
    \multirow{2}{*}{Model} & \multicolumn{2}{c}{\sdss} & \multicolumn{2}{c}{\sqlshare} & \multicolumn{2}{c}{\joinorder}\\
    \cmidrule(lr){2-3} \cmidrule(lr){4-5} \cmidrule(lr){6-7}
     & MAE & HR & MAE & HR & MAE & HR \\
    \midrule
    \gptfour  & \textbf{4.69} & \textbf{0.56} & \textbf{3.96} & \textbf{0.63} & \textbf{3.45} & \textbf{0.57}\\
    \gptthree  & 17.71 & 0.25 & 7.71 & \underline{0.42} & 14.31 & {0.39}\\
    \llama  & \underline{15.60} & 0.33 & \underline{7.57} & {0.40} & 13.11 & {0.39}\\
    \mistralai  & {18.09} & \underline{0.36} & {8.58} & \underline{0.42} & \underline{9.92} & \underline{0.40}\\
    \gemini & 19.78 & 0.34 & 9.79 & 0.38 & 20.22 & 0.32 \\
    \bottomrule  
    \end{tabular}
    \caption{MAE and Hit Rate (HR) for \mtokenl}
    \label{tab:position_prediction_metrics}
\vspace{-4ex}
\end{table}

\paraStart{\mtokenl.} Table~\ref{tab:position_prediction_metrics} compares the performance of various LLMs in predicting the location of the missing token across \sdss, \sqlshare, and \joinorder. The key metrics are Mean Absolute Error (MAE) and Hit Rate (HR), where lower MAE and higher HR indicate better performance.

\gptfour consistently achieves the best results with the lowest MAE and highest HR across all datasets. \llama performs well in \sqlshare but shows weaker results elsewhere. \gptthree and \mistralai provide reasonable performance but with higher MAE and lower HR, reflecting less precision and accuracy.

Most models correctly predict the exact location at least 30\% of the time, except \gptthree in \sdss, where the HR drops to 25\%. Longer queries, especially in \sdss, contribute to higher MAE values, making precise location prediction more difficult.

\takeaway{All models perform better over the missing token tasks than syntax error detection, as missing token identification seems to be a simpler task related to learning frequent patterns. \llama\ shows improved performance due to its broad training in pattern detection. More complex queries tend to increase prediction errors, where complexity is related to different syntactic properties, such as \wcount, \pcount, \nlevel, and \tcount.}


\begin{table}[t!]
    \centering
    \begin{tabular}{lcccc}
    \addlinespace[-1.5ex] 
    \toprule
    Model & Prec. & Rec. & F1  \\
    \midrule
    GPT4       & \textbf{0.88} & \textbf{0.93} & \textbf{0.90} \\
    GPT3.5     & \underline{0.81} & 0.83 & \underline{0.85} \\
    Llama3     & {0.76} & \underline{0.90} & 0.82\\
    MistralAI  & 0.47 & \underline{0.90} & 0.62\\
    Gemini     & 0.71 & {0.73} & 0.72\\
    \bottomrule
    \end{tabular}
    \caption{Acc. for \qperformance}
   \vspace{-2ex}
    \label{tab:runtime_prediction_metrics}
    \vspace{-4ex}
\end{table}

\subsection{Query Performance Prediction}

Table~\ref{tab:runtime_prediction_metrics} shows the performance metrics for the \sdss dataset on the \qperformance task. \gptfour\ achieves the best results, followed by \gptthree\ and \llama, which perform similarly. \mistralai\ and \gemini\ show lower overall performance. Across all models, recall is generally higher than precision, likely due to positive bias. LLMs tend to produce overly optimistic responses, in this case predicting that queries will take longer to run. Additionally, the queries are selected from more complex, lengthy queries in \sdss, which increases the likelihood of being labeled as costly.

\begin{figure}[h]
    \begin{subfigure}[b]{0.22\textwidth}
        \includegraphics[width=\textwidth]{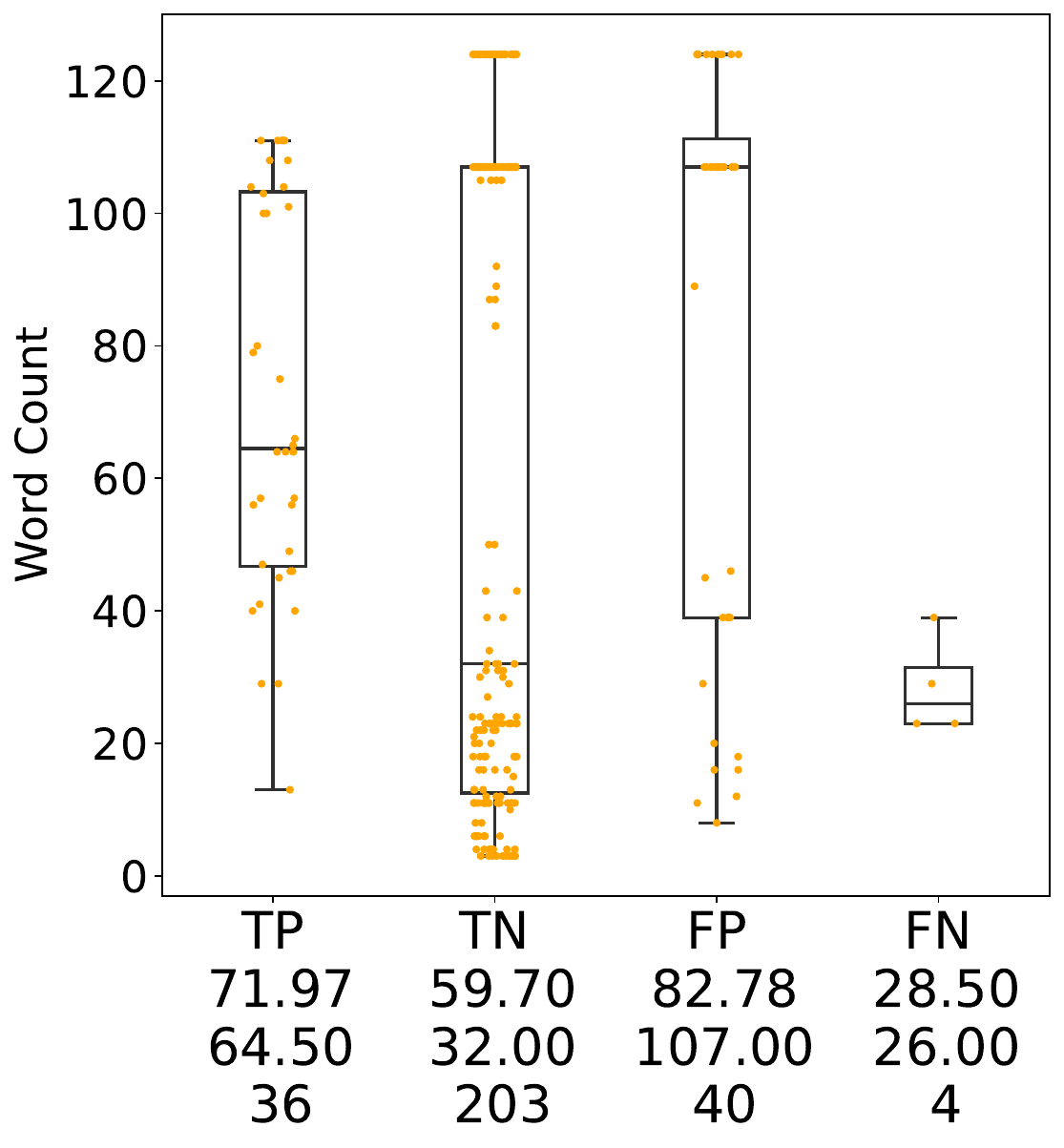}
        \caption{\wcount}
        \label{fig:runtime-wcount}
    \end{subfigure}%
    \begin{subfigure}[b]{0.22\textwidth}
        \includegraphics[width=\textwidth]{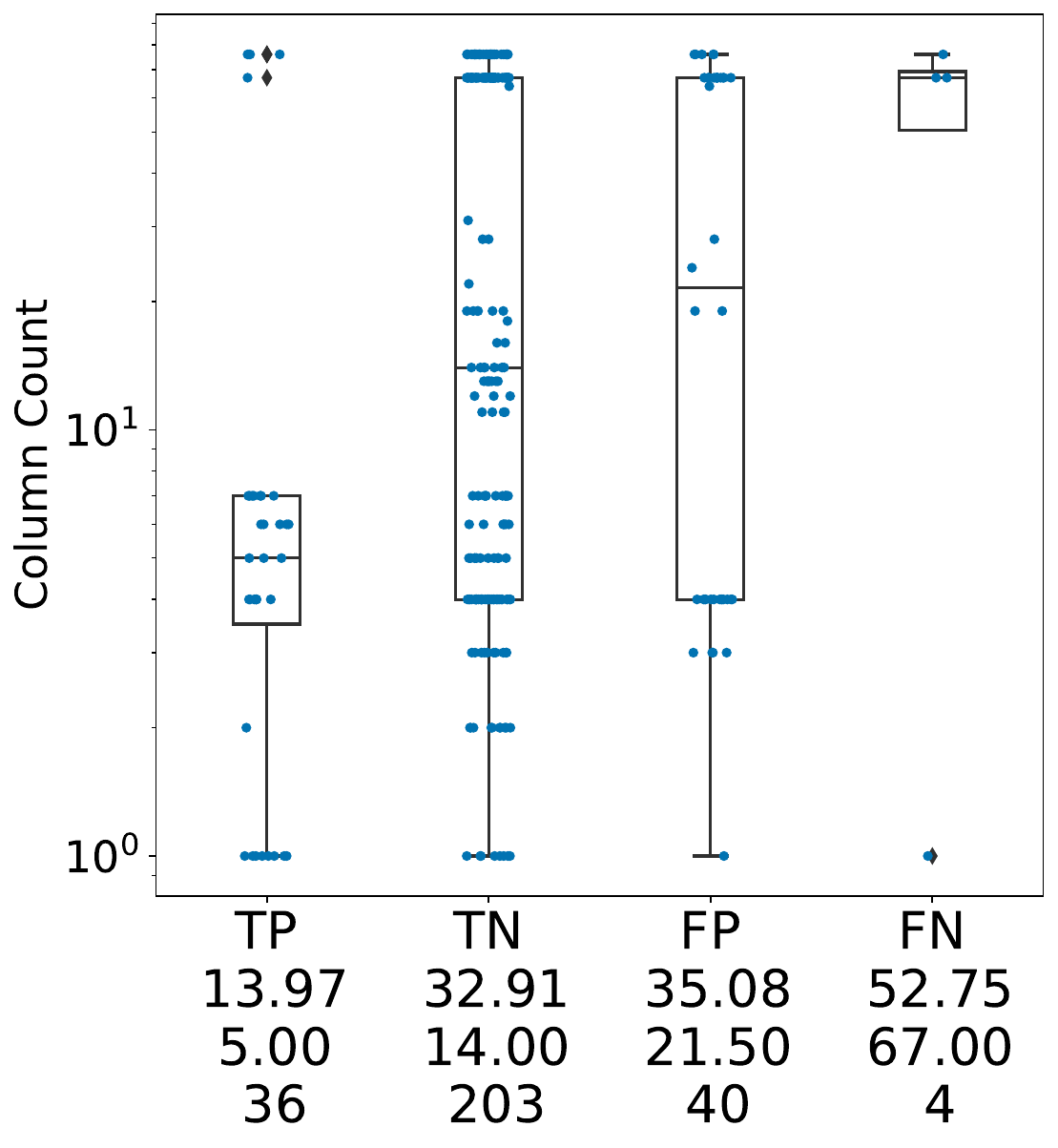}
        \caption{\ccount}
        \label{fig:runtime-ccount}
    \end{subfigure}
    \vspace{-3mm}
    \caption{\mistralai's failure in \qperformance}
    \label{fig:runtime-syntx}
\end{figure}

As with \mtoken, we examined the relationship between syntactic properties and failure rates for this task. The models show strong correlations between \wcount\ and failure, with longer queries leading to more \FP, as shown in Figure~\ref{fig:runtime-wcount} for \mistralai. A similar trend is seen with \ccount\ in Figure~\ref{fig:runtime-ccount}. This suggests that the models mistakenly associate longer queries or those with more columns with higher execution time.

\takeaway{In the query performance prediction task, \gptfour consistently shows the highest accuracy. However, all models tend to overestimate runtimes, leading to higher recall but lower precision, especially for longer and more complex queries. This suggests that improving model training with diverse query types could reduce this bias and enhance prediction accuracy.}


\begin{table}[h!]
    \centering
    \setlength{\tabcolsep}{2pt} 
    \begin{tabular}{lcccccccccc}
    \toprule
    \multirow{3}{*}{\rotatebox[origin=c]{90}{\small Case}} & \multirow{2}{*}{Model} & \multicolumn{3}{c}{\sdss} & \multicolumn{3}{c}{\sqlshare} & \multicolumn{3}{c}{\joinorder}  \\
    \cmidrule(lr){3-5} \cmidrule(lr){6-8} \cmidrule(lr){9-11}
     &  & Prec. & Rec. & F1 & Prec. & Rec. & F1 & Prec. & Rec. & F1 \\
    \midrule
    \multirow{5}{*}{\rotatebox[origin=c]{90}{\small Equivalence}} 
    & \gptfour & \textbf{0.98} & \textbf{1.00} & \textbf{0.99} & \textbf{0.97} & \textbf{1.00} & \textbf{0.99} & \textbf{0.91} & \textbf{1.00} & \textbf{0.95}\\
    & \gptthree  & 0.87 & \underline{0.99} & 0.93 & \underline{0.96} & \textbf{1.00} & \underline{0.98} & {0.83} & \underline{0.99} & {0.90}\\
    & \llama & {0.88} & \textbf{1.00} & {0.93} & {0.94} & 0.98 & 0.96 & \underline{0.87} & \textbf{0.99} & \underline{0.93} \\
    & \mistralai & \underline{0.95} & {0.95} & \underline{0.95} & 0.95 & {0.93} & 0.94 & 0.86 & {0.89} & {0.88} \\
    & \gemini & 0.84 & {0.97} & 0.90 & 0.92 & \underline{0.99} & 0.95 & 0.85 & 0.96 & 0.90 \\
    \midrule
    \multirow{5}{*}{\rotatebox[origin=c]{90}{\small Equiv. Type}} 
    & \gptfour & \textbf{0.99} & \textbf{0.99} & \textbf{0.99} & \textbf{0.98} & \textbf{0.98} & \textbf{0.98} & \textbf{0.95} & \textbf{0.85} & \textbf{0.83} \\
    & \gptthree & \underline{0.97} & \underline{0.91} & \underline{0.91} & \underline{0.96} & \underline{0.92} & \underline{0.94} & {0.90} & {0.78} & {0.77} \\
    & \llama & \underline{0.97} & 0.85 & {0.86} & 0.93 & 0.88 & 0.89 & \underline{0.93} & \underline{0.81} & \underline{0.80} \\
    & \mistralai & {0.85} & {0.76} & {0.80} & {0.92} & {0.88} & {0.89} & {0.84} & {0.68} & {0.68} \\
    & \gemini & 0.86 & 0.72 & 0.71 & 0.91 & 0.85 & 0.87 & {0.87} & {0.77} & {0.75} \\
    \bottomrule
    \end{tabular}
    \caption{Accuracy in \qequiv and \qequivt}
    \label{tab:equiv_metrics}
    \vspace{-6ex}
\end{table}

\subsection{Query Equivalence} \label{sec:equiv-exp}

Table~\ref{tab:equiv_metrics} presents the results for \qequiv (top) and \qequivt (bottom). For both tasks, \gptfour achieves the best performance, with \gptthree and \llama following closely but slightly less consistently. \mistralai and \gemini show more variability and generally lower scores across datasets. A positive bias (higher recall than precision) is noticeable in \qequiv but not in \qequivt, likely due to the latter being a multiclass task rather than binary.

Overall, \qequivt proves more challenging, with lower performance across LLMs and datasets, except for \gptfour, which maintains near-perfect accuracy. This is expected, as determining equivalence is simpler than identifying the type of equivalence. Lower performance in \joinorder and \sdss compared to \sqlshare suggests that longer queries make \qequiv more difficult.

Across all datasets, most LLMs show very few or no \FN, reflected in the high recall. For example, \gptfour records \FP in \sdss (5), \sqlshare (4), and \joinorder (9) but has no \FN. Thus, we focus on \FP to identify where models fail. A common feature of \FP queries is that they involve modified conditions, such as changing values in conditions. For instance, altering ``\textsf{WHERE run = 756 AND field = 103}'' to ``\textsf{WHERE run = 756 AND field = 200}'' or ``\textsf{WHERE run = 756 OR field = 103}.'' This indicates that LLMs struggle with logical reasoning and numerical manipulation, a limitation extensively discussed in the literature~\cite{frieder2024mathematical,Cobbe2021TrainingVT,chowdhery2023palm,wei2022chain,creswellselection,huang2022language} and we confirm in our study. 

\begin{figure}[h]
    \begin{subfigure}[b]{0.22\textwidth}
        \includegraphics[width=\textwidth]{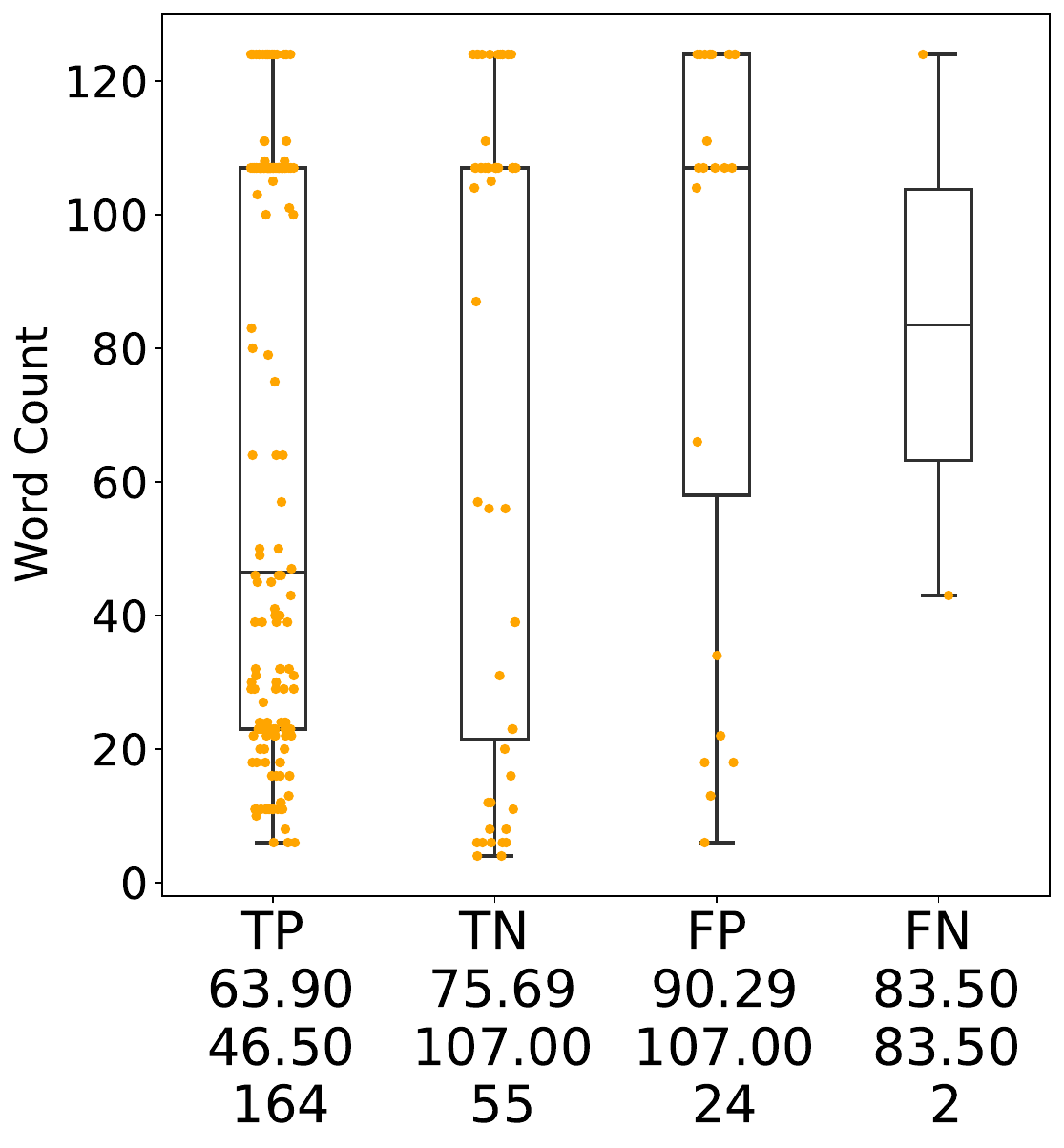}
        \caption{\gptthree in \sdss}
        \label{fig:gpt3.5-qequiv-wcount-sdss}
    \end{subfigure}%
    \begin{subfigure}[b]{0.22\textwidth}
        \includegraphics[width=\textwidth]{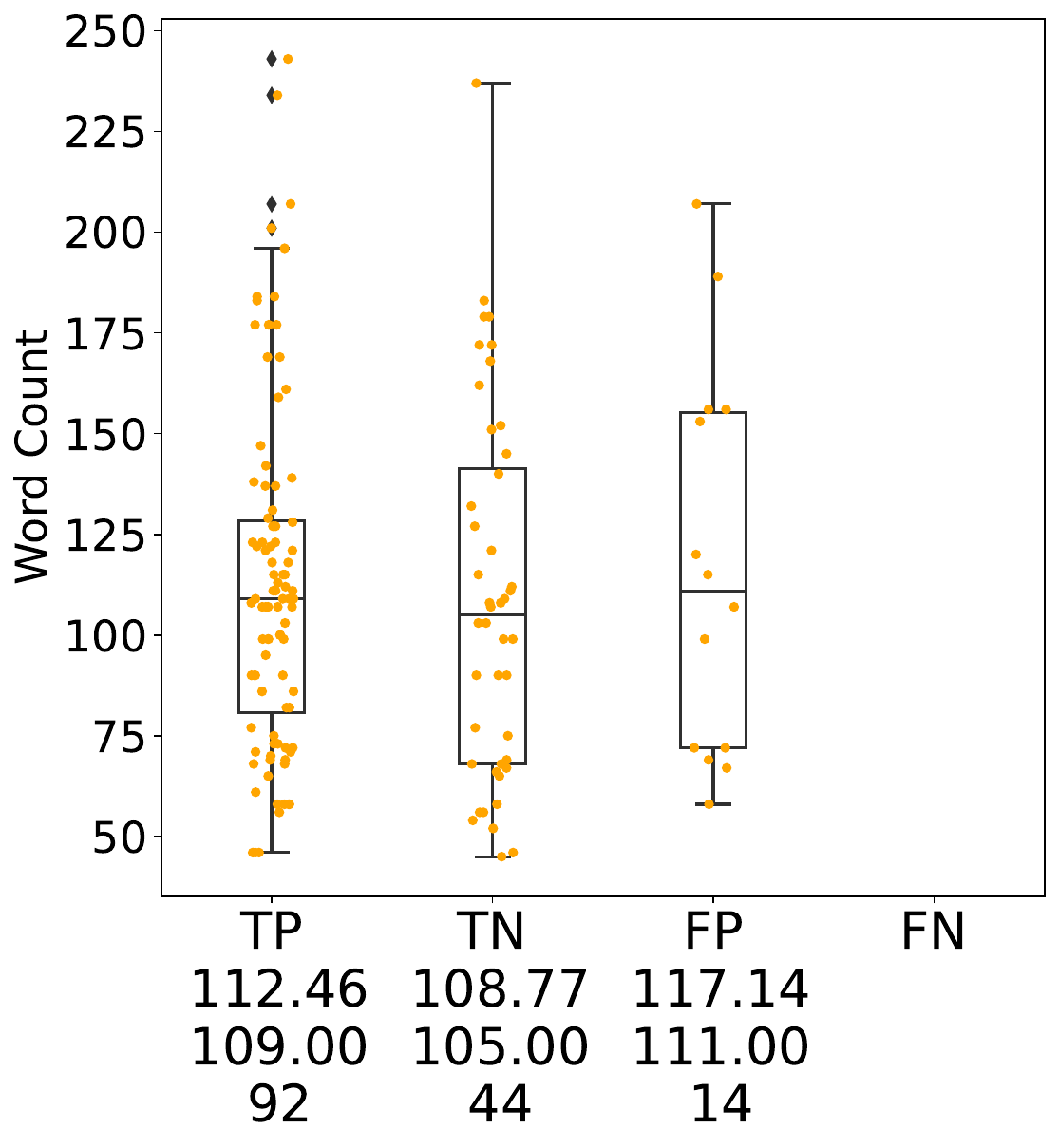}
        \caption{\llama in \joinorder}
        \label{fig:gpt3.5-qequiv-wcount-join}
    \end{subfigure}
    \vspace{-3mm}
    \caption{\wcount and LLM failures in \qequiv.}
    \label{fig:qequiv-wcount}
\end{figure}

\begin{figure}[h]
    \begin{subfigure}[b]{0.22\textwidth}
        \includegraphics[width=\textwidth]{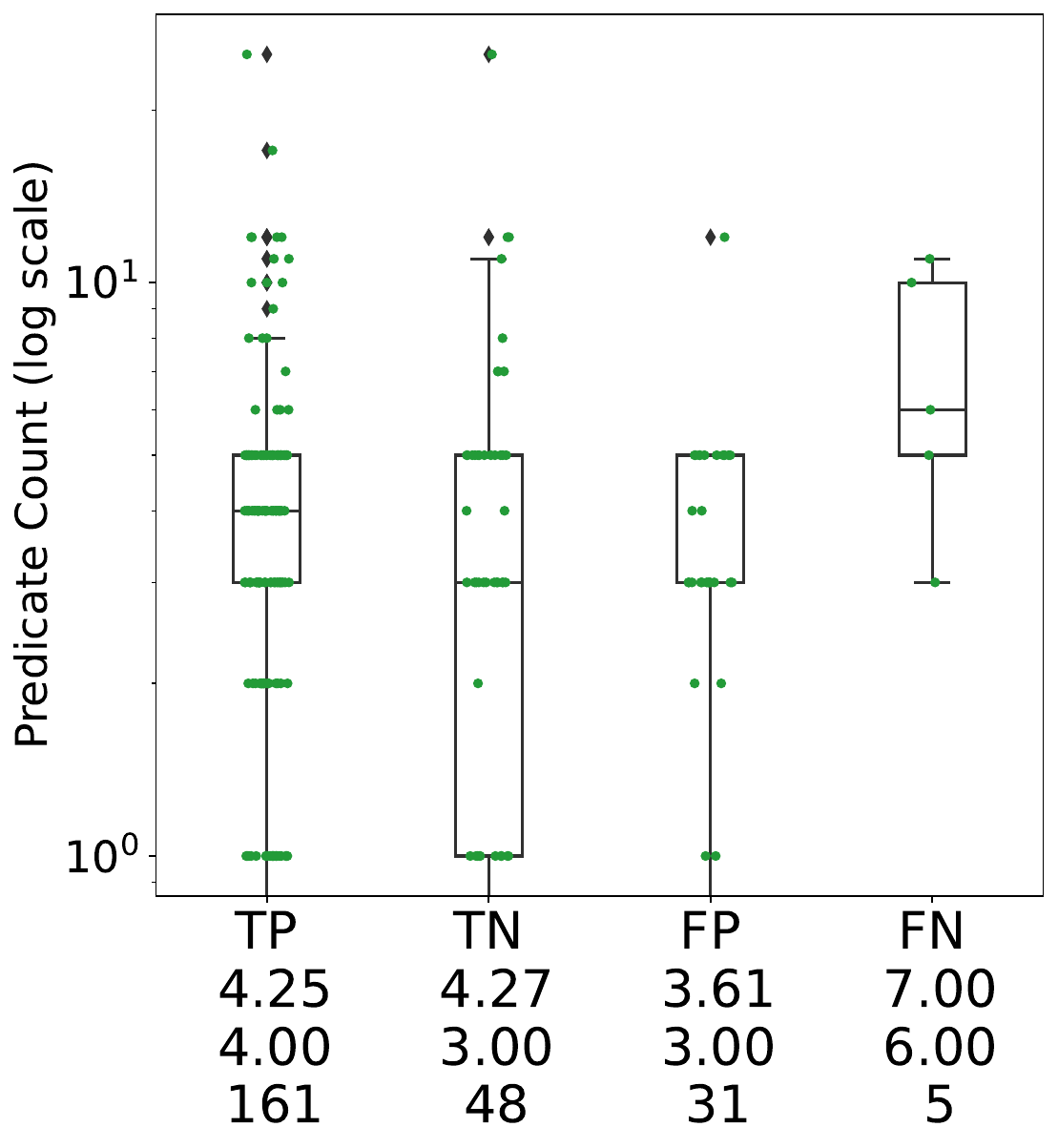}
        \caption{\gemini in \sdss}
        \label{fig:gemini-qequiv-pcount-sdss}
    \end{subfigure}%
    \begin{subfigure}[b]{0.22\textwidth}
        \includegraphics[width=\textwidth]{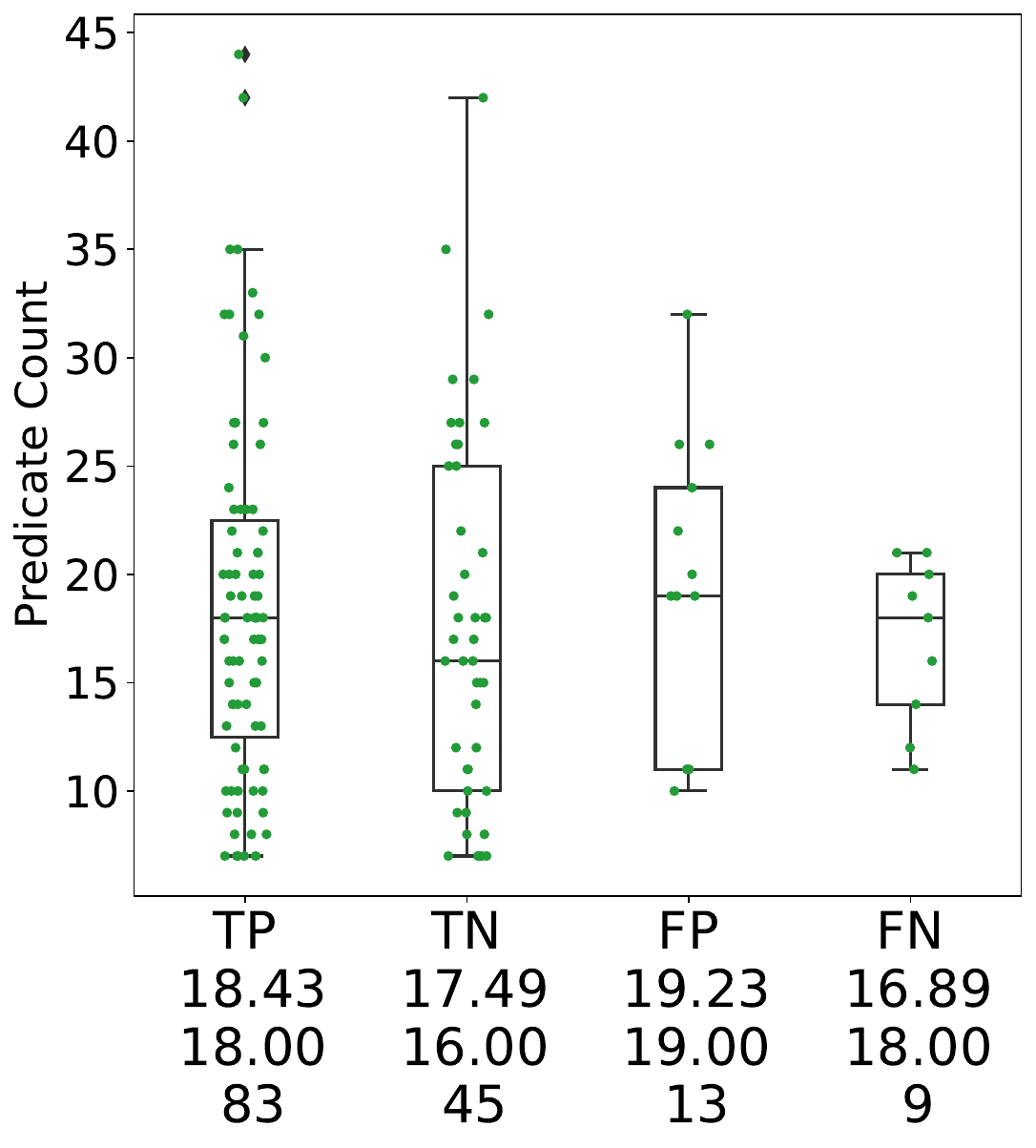}
        \caption{\mistralai in \joinorder}
        \label{fig:mistralai-qequiv-pcount-join}
    \end{subfigure}
    \vspace{-3mm}
    \caption{\pcount and LLM failure in \qequiv.}
    \label{fig:qequiv-pcount}
\end{figure}

In addition to logical reasoning and numerical issues, these problems become more pronounced in more complex queries, such as those with longer lengths or more tables and predicates. For \gptfour in \sdss, all 5 \FP involve queries over 100 words, a pattern also observed in \gptthree (Figure~\ref{fig:qequiv-wcount}). In \joinorder, where most queries are lengthy, both \FP and \FN occur more frequently across all LLMs. We report only \llama for \joinorder, as other LLMs exhibit similar trends. Considering \tcount as a complexity parameter, in \joinorder, all \FP occur in queries with more than 8 tables. Figure~\ref{fig:qequiv-pcount} shows that in \sdss, \FP occurs in queries with 5 or more predicates. Similarly, in \joinorder (Figure~\ref{fig:qequiv-pcount}), all \FP across models are caused by queries with over 19 predicates. We include the figure for \mistralai, as all LLMs exhibit the same pattern. These suggest that \qequiv and \qequivt are more difficult for complex queries (longer queries with more predicates and tables).


\takeaway{In the query equivalence task, \gptfour performs best across datasets. However, distinguishing between different types of equivalence (\qequivt) proves more difficult, especially for \gemini and \mistralai. The errors mainly stem from challenges in understanding complex query conditions, highlighting the need for better SQL logic comprehension in LLMs.}

\subsection{Case Study: Query Explanation} \label{sec:explanation}

We study the \qexplain task and analyze several cases where LLMs failed to provide accurate explanations for SQL queries. The queries are presented in Listing~\ref{query:explanations}. Here we present the correct ground truth descriptions from \spider, the erroneous explanations generated by the models, and briefly provide our analysis of the case:

\paraStart{Q15.} The query finds the number of students who participate in the tryout for each college, ordered by descending count. \gemini incorrectly describes the query as: ``Counts the occurrences of each unique value in the \texttt{cName} column.'' This explanation reduces the query to a simple counting of values in the \texttt{cName} column, ignoring the fact that the query specifically searches for students in tryouts, which is essential to convey its full meaning. 

\paraStart{Q16.} The query identifies the maximum number of times a course enrollment result can appear in different transcripts and displays the course enrollment ID. \gemini's explanation is: ``Finds the student course ID with the highest number of occurrences.'' While this partially captures the query’s purpose, it misses the query context of searching in transcripts. 

\paraStart{Q17.} The query finds the name and location of stadiums where concerts took place in both 2014 and 2015. \gptfour explains it as: ``The query identifies stadiums that hosted concerts in both 2014 and 2015.'' This only partially explains the query and does not include the selected attributes. This issue occurs because LLMs often focus on capturing the overall semantics of a query but overlook specific details, such as selected attributes, especially in more complex tasks.



\paraStart{Q18.} The query retrieves the number of cylinders for the Volvo car with the least acceleration. \llama incorrectly explains: ``This SQL query retrieves the number of cylinders of the Volvo car with the fastest acceleration.'' The models misinterpret the ``\texttt{ORDER BY ... ASC LIMIT 1}'' clause, misunderstanding that the query is looking for the slowest car (lowest acceleration) rather than the fastest. Only \mistralai correctly explains this query.

\takeaway{These examples highlight a common issue with LLMs when explaining SQL queries: they often miss or misinterpret key details, particularly in tasks requiring context retention. While models may capture parts of a query, they frequently fail to provide complete and accurate explanations. This reflects known limitations of LLMs in retaining context and applying knowledge to specific scenarios~\cite{shinn2023contextforget,liu2023attention,petroni2020contextforget}.}


\begin{lstlisting}[label={query:explanations},caption={Query statements with inaccurate explanations}]
-- Q15:
SELECT count(*),cName FROM tryout 
GROUP BY cName ORDER BY count(*) DESC
-- Q16:
SELECT count(*),student_course_id FROM Transcript_Cnt 
GROUP BY student_course_id ORDER BY count(*) DESC LIMIT 1
-- Q17:
SELECT S.name,S.loc 
FROM concert AS C JOIN stadium AS S 
ON C.stadium_id = S.stadium_id WHERE C.Year = 2014 
INTERSECT 
SELECT S.name,S.loc FROM concert AS C JOIN stadium AS S
ON C.stadium_id = S.stadium_id WHERE C.Year = 2015
-- Q18:
SELECT C.cylinders FROM CARS_DATA AS C 
JOIN CAR_NAMES AS T ON C.Id = T.MakeId 
WHERE T.Model = 'volvo' 
ORDER BY C.accelerate ASC LIMIT 1;
\end{lstlisting}

\subsection{Reflections on SQL Understanding}\label{sec:reflection}
We designed our SQL tasks to probe basic skills in understanding (as described in Section~\ref{sec:intro}): recognition, semantics, context, and coherence.

\paraStart{Demonstrated Skills.} Our results show that LLMs, particularly GPT-4, perform well in tasks requiring recognition and context. For instance, in syntax error detection, the models showed high precision in recognizing violations of SQL syntax, indicating a strong ability to identify and interpret the structure of SQL queries. Similarly, missing token identification rely on the model's context-awareness, as the ability to predict missing elements in a query depend on understanding the surrounding tokens and their relationships. This success highlights that LLMs effectively interpret the context within a query to identify missing or incorrect components.

\paraStart{Limitations and Shortcomings.} \eat{However, the models also revealed limitations, particularly in} 
The models were less successful at tasks requiring deeper coherence and semantic understanding. For example, query equivalence tasks \eat{which require recognizing when two syntactically different queries produce the same result,} proved to be challenging, especially for longer and more complex queries. This difficulty suggests that while LLMs understand surface-level structures, they struggle with deeper semantic coherence, and logical connections within queries.  For query performance estimation, the models often overestimated runtimes, indicating that they lack a nuanced understanding of how query complexity, and database-specific factors interact to affect performance.

These observations suggest that while LLMs are adept at recognizing patterns and context, their ability to fully ``understand" the deeper semantics and logical coherence of SQL queries is still developing. Future work should focus on refining these models to address these shortcomings, improving their overall SQL proficiency.

\section{Related Work} \label{sec:related-work}

Recent advancements in LLMs have led to innovative approaches in data management, tackling tasks such as data wrangling, entity matching, table manipulation, and text-to-SQL generation.

Li et al.~\cite{li2024towards} propose an LLM-based approach for data wrangling that leverages code generation for structured data transformations. This method significantly reduces computational costs compared to row-by-row processing, which is common in traditional LLM approaches. Their work highlights the importance of deterministic transformations to enhance model interpretability and reliability for data tasks like unit conversion and error detection.

In entity matching, Zhang et al.~\cite{zhang2024anymatch} introduce AnyMatch, a zero-shot entity matching model that achieves competitive performance using a small, specialized LLM. By utilizing efficient data selection techniques, this model performs comparably to larger models like GPT-4, while requiring fewer computational resources. Complementing this, Steiner et al.~\cite{steiner2024fine} explore the benefits of fine-tuning LLMs for entity matching, showing significant performance improvements but also noting that fine-tuning may reduce cross-domain generalization.

LLMs have been used for table manipulation as shown in Li et al.~\cite{li2024table} with Table-GPT. This fine-tuned model is designed for tasks such as data cleaning and table-based question answering. The study demonstrates that LLMs trained on natural language text face limitations when handling two-dimensional tabular data, and that table-specific fine-tuning is necessary to overcome these challenges.

LLMs have also made significant progress in text-to-SQL tasks. Surveys by Hong et al.~\cite{hong2024next} and Gao et al.~\cite{gaotext} provide overviews of how LLMs handle complex and cross-domain SQL generation. These studies highlight that while LLMs perform well on simpler queries, their accuracy drops with more complex structures involving nested queries, joins, and aggregations.

These works underscore the expanding role of LLMs in data management, from entity matching and data wrangling to text-to-SQL. Despite their potential for automating complex tasks, further research is needed to overcome challenges in efficiency, scalability, and generalization across domains.

\section{Conclusion and Future Work}\label{sec:future}

In this paper, we study the proficiency of state-of-the-art LLMs towards ``understanding" SQL.  We evaluate their performance on key SQL tasks such as syntax error identification, missing token identification, query equivalence, query performance estimation, and query explanation. Our evaluation revealed that all models perform well on tasks requiring recognition and context.  \gptfour consistently outperformed other models, particularly in handling complex SQL queries, while \gptthree, \mistralai, and \llama showed strong capabilities in pattern recognition. In contrast, Gemini struggled with all SQL-specific tasks, particularly error detection. Despite these strengths, LLMs faced challenges with long and complex queries, an inability to pinpoint the exact location of missing tokens, and struggled with tasks requiring semantic coherence and logical connections within queries.  As next steps, we will explore fine-tuning to handle query complexity and dynamic prompt tuning (to improve accuracy), and barriers to using LLMs for query recommendation and query optimization.
We anticipate that targeted fine-tuning and dynamic prompt adjustment could significantly mitigate current limitations in handling complex queries and improve task-specific performance. This enhancement of LLMs is expected to bridge the gap between AI capabilities and real-world SQL needs, enabling better integration with database systems.


\clearpage

\bibliographystyle{ACM-Reference-Format}
\bibliography{short-ref}


\end{document}